\documentclass[]{pasj02} 
\usepackage[switch,mathlines]{lineno} 
\usepackage{color}
\usepackage{graphicx}
\usepackage{url}
\usepackage[markup=default]{changes}
\usepackage{ulem}
\definecolor{Changes@AddedColor}{named}{blue}
\definecolor{Changes@DeletedColor}{named}{red}
\definecolor{Changes@ReplacedColor}{named}{magenta}
\jyear{2025}
\Received{}
\Accepted{}

\begin{document} 

\title{
A Possible Shutting-Down Event of Mass Accretion in An Active Galactic Nucleus at $z\sim\roughzclq$
}

\author{
Tomoki \textsc{Morokuma},\altaffilmark{1}\altemailmark\orcid{0000-0001-7449-4814}\email{tomoki.morokuma@p.chibakoudai.jp}
Malte \textsc{Schramm},\altaffilmark{2}\orcid{0000-0001-7825-0075}
Toshihiro \textsc{Kawaguchi},\altaffilmark{3}\orcid{0000-0002-3866-9645}
Josefa Becerra \textsc{Gonz\'{a}lez},\altaffilmark{4,5}\altemailmark\orcid{0000-0002-6729-9022}
Jose Antonio \textsc{Acosta-Pulido},\altaffilmark{4,5}\altemailmark\orcid{0000-0002-0433-9656}
Nieves \textsc{Castro-Rodr\'{i}guez},\altaffilmark{6,4,5}\altemailmark\orcid{0000-0002-4269-0279}
Kana \textsc{Morokuma-Matsui},\altaffilmark{7}\altemailmark\orcid{0000-0003-3932-0952}
Shintaro \textsc{Koshida},\altaffilmark{8}
Junko \textsc{Furusawa},\altaffilmark{9}
Hisanori \textsc{Furusawa},\altaffilmark{9}\orcid{0000-0002-6174-8165}
Tsuyoshi \textsc{Terai},\altaffilmark{8}\orcid{0000-0003-4143-4246}
Fumi \textsc{Yoshida},\altaffilmark{10,11}\orcid{0000-0002-3286-911X}
Kotaro \textsc{Niinuma},\altaffilmark{12}\orcid{0000-0002-8169-3579}
Yoshiki \textsc{Toba},\altaffilmark{13,9,14,15}\orcid{0000-0002-3531-7863}
}
\altaffiltext{1}{Astronomy Research Center, Chiba Institute of Technology, 2-17-1, Tsudanuma, Narashino, Chiba 275-0016, Japan}
\altaffiltext{2}{Universit\"{a}t Potsdam, Karl-Liebknecht-Str. 24/25, D-14476 Potsdam, Germany}
\altaffiltext{3}{Graduate School of Science and Engineering, University of Toyama, Gofuku 3190, Toyama 930-8555, Japan}
\altaffiltext{4}{Instituto de Astrofisica de Canarias E-38200 La Laguna, Tenerife, Spain}
\altaffiltext{5}{Universidad de La Laguna, Departamento de Astrofisica, E-38206 La Laguna, Tenerife, Spain}
\altaffiltext{6}{GRANTECAN, Cuesta de San Jos\'{e} s/n, E-38712, Bre\~{n}a Baja, La Palma, Spain}
\altaffiltext{7}{Institute of Astronomy, Graduate School of Science, University of Tokyo, 2-21-1, Osawa, Mitaka, Tokyo 181-0015, Japan}
\altaffiltext{8}{Subaru Telescope, National Astronomical Observatory of Japan, National Institutes of Natural Sciences, Hilo, HI 96720, USA}
\altaffiltext{9}{National Astronomical Observatory of Japan, 2-21-1 Osawa, Mitaka, Tokyo 181-8588, Japan}
\altaffiltext{10}{School of Medicine, Department of Basic Sciences, University of Occupational and Environmental Health, 1-1 Iseigaoka, Yahata, Kitakyusyu, Fukuoka 807-8555, Japan}
\altaffiltext{11}{Planetary Exploration Research Center, Chiba Institute of Technology, 2-17-1, Tsudanuma, Narashino, Chiba 275-0016, Japan}
\altaffiltext{12}{Graduate School of Sciences and Technology for Innovation, Yamaguchi University, 1677-1, Yoshida, Yamaguchi 753-8512, Japan}
\altaffiltext{13}{Department of Physical Sciences, Ritsumeikan University, 1-1-1, Noji-higashi, Kusatsu, Shiga 525-8577, Japan}
\altaffiltext{14}{Academia Sinica Institute of Astronomy and Astrophysics, 11F of Astronomy-Mathematics Building, AS/NTU, No.1, Section 4, 14Roosevelt Road, Taipei 10617, Taiwan}
\altaffiltext{15}{Research Center for Space and Cosmic Evolution, Ehime University, 2-5 Bunkyo-cho, Matsuyama, Ehime 790-8577, Japan}



\KeyWords{
galaxies: active --- 
quasars: individual (SDSS~J021801.90-003657.7) --- 
quasars: emission lines --- 
accretion, accretion disks --- 
surveys
}

\maketitle

\newcommand{\declinefactorobservednew}{20-30}
\newcommand{\declinefactoragnnew}{50}
\newcommand{\declinefactoreddratenew}{\declinefactoragnnew}
\newcommand{\releddratefirst}{1.000}
\newcommand{\releddratelast}{0.020}
\newcommand{\avvaluelast}{2.2}
\newcommand{\declinefactoreddrate}{50}
\newcommand{\declinefactorsdssimgtospecnew}{8}
\newcommand{\declinefactorsdssimgtospec}{8}
\newcommand{\bhmass}{$M_{\rm{BH}}$}
\newcommand{\bhmassfiducialcorrectedforfading}{$8.6-8.9$}
\newcommand{\correctedforfading}{3}
\newcommand{\bhmassfiducial}{$1.8\times10^{8}$}
\newcommand{\bhmasscfour}{$1.2\times10^{8}$}
\newcommand{\logbhmassmgtworakshit}{$8.25\pm0.42$}
\newcommand{\logbhmassmgtwowu}{$8.44\pm0.21$}
\newcommand{\logbhmasscfourrakshit}{$8.09\pm0.07$}
\newcommand{\logbhmasscfourwu}{$8.23\pm0.11$}
\newcommand{\solarmass}{$M_\odot$}
\newcommand{\rathisobj}{\timeform{2h18m01.90s}}
\newcommand{\decthisobj}{\timeform{-0D36'57.8''}}
\newcommand{\numofquasarswithinhscfootprints}{31,549}
\newcommand{\roughnumofquasarswithinhscfootprints}{$3\times10^{4}$}
\newcommand{\redshiftqso}{$1.767$}
\newcommand{\magvarithre}{0.5}
\newcommand{\cmodelminuspsfthre}{0.15}
\newcommand{\sdssmagg}{20.62}
\newcommand{\sdssmagr}{20.72}
\newcommand{\sdssmagi}{20.55}
\newcommand{\sdssmagerrg}{0.02}
\newcommand{\sdssmagerrr}{0.03}
\newcommand{\sdssmagerri}{0.04}
\newcommand{\hscmagg}{23.87}
\newcommand{\hscmagr}{23.24}
\newcommand{\hscmagi}{23.28}
\newcommand{\hscmagerrg}{0.02}
\newcommand{\hscmagerrr}{0.01}
\newcommand{\hscmagerri}{0.04}
\newcommand{\mjdsdssspec}{57016}
\newcommand{\mjdlrisspec}{59791}
\newcommand{\sdsstohscmagvarig}{3.26}
\newcommand{\sdsstohscmagvarir}{2.52}
\newcommand{\sdsstohscmagvarii}{2.73}
\newcommand{\sdsstohscmagvarierrorg}{0.03}
\newcommand{\sdsstohscmagvarierrorr}{0.03}
\newcommand{\sdsstohscmagvarierrori}{0.06}
\newcommand{\separcsecnearbygalaxy}{1.71}
\newcommand{\tinfl}{$t_{\rm{infl}}$}
\newcommand{\tdustred}{$t_{\rm{dust}}$}
\newcommand{\eddratiohighnewimg}{0.4}
\newcommand{\eddratiohighnewsdssspec}{0.06}
\newcommand{\eddratiohighnew}{0.4}
\newcommand{\eddratiolownew}{0.008}
\newcommand{\palomartelescope}{the 48-inch Palomar Oschin Schmidt telescope}
\newcommand{\redposs}{{\it Red}}
\newcommand{\blueposs}{{\it Blue}}
\newcommand{\irposs}{{\it IR}}
\newcommand{\stellarmass}{$1.4\times10^{11}$}
\newcommand{\orgstellarmass}{$2.7\times10^{11}$}
\newcommand{\sfrate}{zero}
\newcommand{\kmpersec}{km~s$^{-1}$}
\newcommand{\fluxunit}{erg~s$^{-1}$~cm$^{-2}$}
\newcommand{\zclq}{1.767}
\newcommand{\roughzclq}{1.8}


\begin{abstract}
We present the discovery of a large gradual apparent fading event 
in optical and near-infrared wavelengths in a quasar at $z=\zclq$ 
by a factor of 
$\sim\declinefactorobservednew$ (in optical)
over a period of
$\sim20$~years in the observed frame. 
This pronounced
fading trend in brightness was first identified 
by comparing the magnitudes measured 
in the Subaru/Hyper Suprime-Cam (HSC) images 
with those in the Sloan Digital Sky Survey (SDSS) images 
for $\sim$\roughnumofquasarswithinhscfootprints\  
quasars spectroscopically identified by SDSS. 
We performed follow-up observations, including 
optical imaging and spectroscopy as well as 
near-infrared imaging, 
with $>4$m-class telescopes 
such as Subaru, GTC, Keck, and SOAR telescopes. 
We combine these new data with the archival data 
to examine the variability behavior 
over $\sim$20~years in detail and 
even the longer-term trend of the variability over $\sim70$~years 
in the observed frame. 
We find that 
(i) the AGN component likely faded
by a factor of $\sim\declinefactoragnnew$ 
from the early 2000s to 2023 
and 
(ii) the observed brightness decline is best explained by 
a substantial decrease in accretion rate rather than 
time-varying line-of-sight dust obscuration. 
These findings are derived 
from multi-component (time-varying AGN + constant galaxy) spectral 
energy distribution fitting over multi-epochs, 
which is well consistent with the optical spectra. 
The Eddington ratio decreases 
by a factor of 
$\sim\declinefactoragnnew$, 
from $\sim\eddratiohighnew$ to $\sim\eddratiolownew$ 
if we use the black hole mass measured with the SDSS spectrum, 
which could be highly uncertain because of the very large variability. 
The total brightness is dominated by the host galaxy 
in the rest-frame optical wavelength 
rather than the AGN as of 2023. 
\end{abstract}


\section{Introduction}\label{sec:secinpaper_intro}


The cosmological growth of supermassive black holes 
(SMBHs) in galaxies has long been of great interest. 
As discussed in the so-called So\l{}tan argument \citep{Soltan:1982vm} 
and subsequent observational and theoretical evidence, 
the lifetime of active galactic nuclei (AGNs) 
is shorter than the age of the universe 
and is estimated to be as short as 
an order of $10^{8}$~yr \citep{Marconi:2004tf}, 
$\sim10^{6-8}$~yr \citep{Hopkins2006ApJS..163....1H}, 
and 
$\sim10^5$~yr \citep{Schawinski2015MNRAS.451.2517S}. 
Given the short AGN lifetime 
and 
ubiquity of SMBHs at galaxy centers, 
mass accretion activity observed as an AGN 
starts at some time and ends at some time in the cosmological history. 
Understanding what triggers the onset and cessation of accretion 
is, therefore, a key goal.

The number of AGNs identified in many wide-field spectroscopic surveys 
has been dramatically increasing, 
such as 
the 2dF Galaxy Redshift Survey (2dFGRS; \cite{Colless:2001aa}), 
the 6dF Galaxy Survey (6dFGS; \cite{Jones:2004aa}), 
the Sloan Digital Sky Survey 
(SDSS; \cite{york2000}), 
as follow-up observations based on wide-field optical imaging 
and other wavelength data. 
The latest quasar catalogs contain 
several hundred thousand 
quasars \citep{Veron-Cetty:2010aa,paris2018,Lyke:2020aa}. 
These quasars have been 
automatically monitored mainly in optical wavelengths 
in numerous imaging survey projects conducted 
since the discovery of 
the accelerating expansion of the Universe 
based on 
Type Ia supernova observations \citep{Riess:1998aa,Perlmutter:1999aa}. 
Monitoring hundreds of thousands of quasars over decades is, in a rough sense, 
equivalent to observing a single quasar for several million years. 
This opens up the possibility of detecting rare events that unfold
on such long timescales. 

Such rare events in AGNs include 
drastic change in mass accretion rate 
of 
SMBHs, sometimes accompanied by change of 
accretion state in an accretion disk. 
Recent studies indicate that Eddington ratios of 
AGNs showing drastic changes 
are as low as $<0.1$ or 0.02 \citep{macleod2019,Noda:2018aa}. 
There are different types of accretion modes in different AGNs, 
including standard accretion \citep{shakura1973} 
and 
optically-thin 
radiatively inefficient accretion flow (RIAF; \cite{Narayan:1994aa,Narayan:1995ab}) 
or 
advection-dominated accretion flow (ADAF; \cite{Yuan:2014aa}). 
Switching these 
various states to another state 
have been observed for low-mass black hole systems 
such as X-ray binaries \citep{Remillard:2006aa}, but, 
it remains unclear for larger black holes, i.e., SMBHs in AGNs. 

Turning-on/off phenomena of mass accretion are 
expected to show 
large-amplitude brightness variability 
\citep{Schawinski2015MNRAS.451.2517S}. 
A subpopulation of AGN, as observed to be blazars, 
have been known to exhibit large variability 
but 
due to beaming effects of relativistic jets 
\citep{Ulrich:1997aa}, not directly related to turning-on/off events. 
But, recently 
a part of AGNs without evidence of the existence of jets 
are found to show dramatic temporal changes in 
rest-frame UV-optical continuum luminosity 
\citep{Kelly:2009wq,MacLeod:2010vd,Rumbaugh:2018tm,Dexter:2019ab}. 
These phenomena are interpreted as 
AGN type transition (from obscured population to unobscured population, 
and vice versa) or turn-on/off of central mass accretion activity. 
Some of such AGNs show 
the temporal change of X-ray hardness indicating 
the temporal change of the line-of-sight obscuring material 
\citep{piconcelli2007,lamassa2015}. 
Another group of such AGNs show 
(dis)appearance of broad emission lines 
\citep{lamassa2015,macleod2016,ruan2016,wang2018,macleod2018,stern2018,Wang:2019aa,Ross:2020aa,cooke2020,Nagoshi:2021aa,potts2021,Jiang:2021aa,Wang:2022aa,green2022,Guo:2024aa,Guo:2024ab}   
indicating the change of the accretion state, 
while some of them only show large brightness change 
without clear evidence for change in either accretion state or line-of-sight obscuration. 

These dramatic changes have been 
found and studied in great detail 
for nearby or low-redshift sources, 
for example, 
Mrk~590 \citep{mathur2018} 
and 
NGC~5548 \citep{dehghanian2019}. 
In Arp~187, 
\citet{ichikawa2019} found 
oxygen forbidden line flux excess compared to central X-ray flux obtained with 
Nuclear Spectroscopic Telescope Array (NuSTAR; \cite{Harrison:2013aa}), 
interpreted as a quick death of the quasar. 
A subsequent study by \citet{Pflugradt:2022aa} also showed 
a population of declining AGNs with [O\,{\sc iii}] flux excess 
compared to mid-infrared (MIR) flux. 
The origin of rapid X-ray change with NuSTAR data 
is also intensively discussed for NGC~3627 
\citep{Saade:2022aa,Esparza-Arredondo:2020aa} and other nearby galaxies \citep{Saade:2022aa}. 
Systematic searches for AGNs showing dramatic change have been also done 
with Mapping Nearby Galaxies at Apache Point Observatory (MaNGA) survey data \citep{Hon:2020vb}, 
SkyMapper \citep{hon2022} and 
Pan-STARRS (PS1) data \citep{Senarath:2021aa}. 

Especially for AGNs with large brightness decline, 
the brightness contrast of the host galaxies relative to AGNs 
is dramatically improved to study the host galaxy properties 
in great detail 
\citep{charlton2019,Jin:2022ud}. 
In terms of general views of AGN host galaxies, 
such fading AGNs provide us 
with a unique opportunity 
to closely examine their central regions 
of AGN host galaxies \citep{Dodd:2021aa}. 

In this paper, we use the photometric catalogs 
from the SDSS (MJD$\sim$52,000-55,000~days) 
and Hyper Suprime-Cam (HSC; \cite{Miyazaki:2018aa}; 
on the 8.2-m Subaru Telescope) data 
for the spectroscopically identified 
quasars \citep{paris2018} 
to search for largely fading quasars between these two datasets. 
For the HSC data, we use those taken 
in the Strategic Survey Program (SSP) for HSC (HSC-SSP; \cite{aihara2018}) 
where the SDSS data is available for the entire survey region. 
We here focus on one extreme quasar, showing the largest decline 
between the SDSS and HSC imaging data; 
SDSS~J021801.90-003657.7 at $z=\zclq$ 
(hereafter, J0218-0036). 

The paper structure is as follows. 
We describe the discovery of the large and rapid fading of a quasar, 
J0218-0036, in \S~\ref{sec:secinpaper_target}. 
We also summarize follow-up observations and archival data of the quasar. 
We discuss the temporal change of the spectral energy distributions (SEDs) 
in \S\ref{sec:secinpaper_seddecompose} 
and variability timescales in \S\ref{sec:secinpaper_timescalevari}. 
In \S\ref{sec:secinpaper_summary}, we summarize the contents of this paper. 
All the magnitudes are measured in the AB system. 
Cosmological parameters used in this paper are 
$H_0=70$ [km~s$^{-1}$~Mpc$^{-1}$], 
$\Omega_M=0.3$, and $\Omega_\Lambda=0.7$.

\section{Large and rapid fading of the quasar J0218-0036}\label{sec:secinpaper_target}

\begin{figure*}[!htbp]
 \begin{center}
    \includegraphics[angle=0,width=179.5mm,bb=0 0 1920 944]{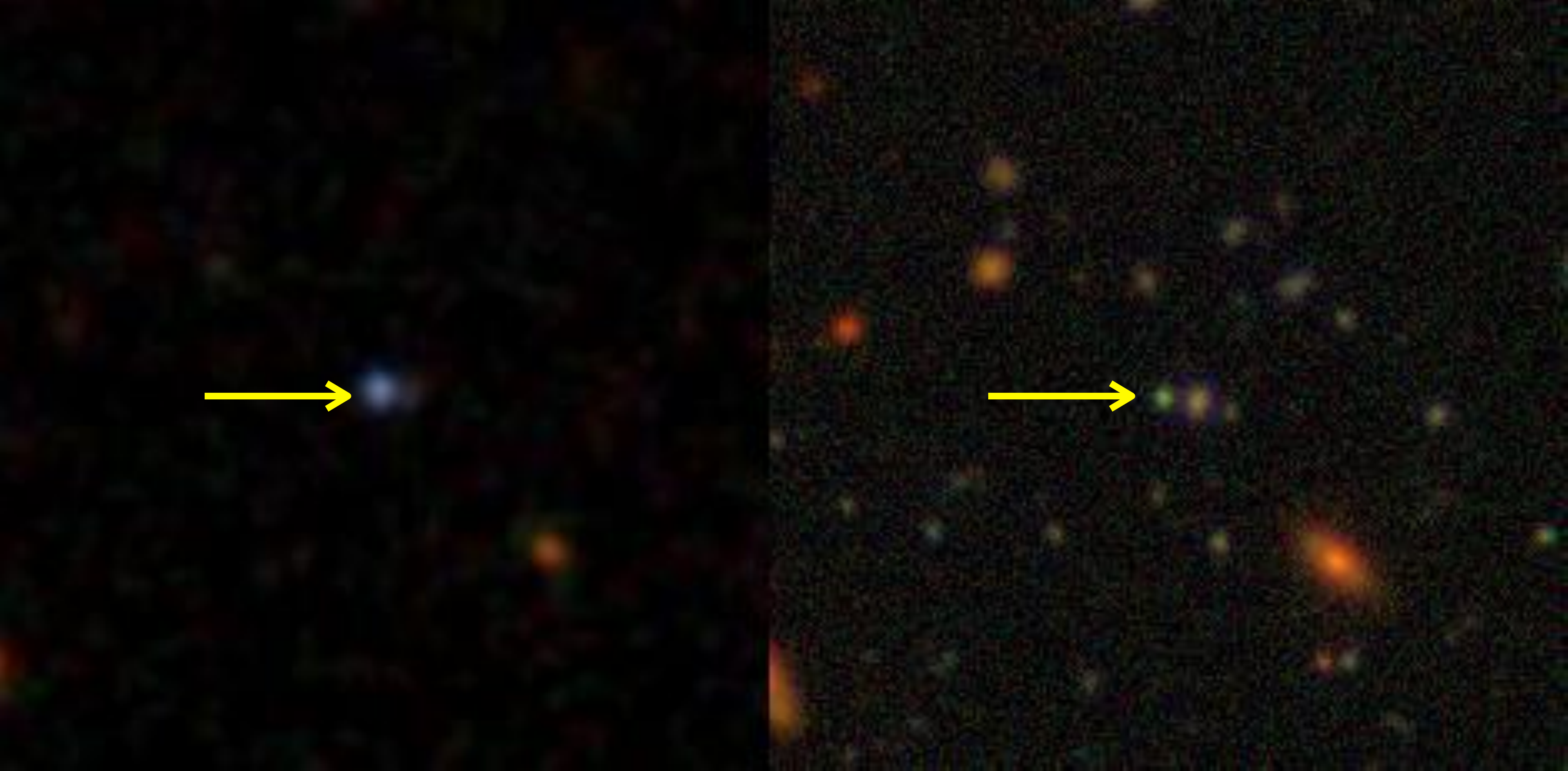}
 \end{center}
\caption{Optical 3-color ($gri$) SDSS and HSC images (hscMap) 
of J0218-0036 field in the left and right panels, respectively. 
The central object is 
the quasar J0218-0036 at $z=\zclq$. 
The depths are greatly different and 
the clear fading of the quasar can be seen 
by its transition from a blue bright point to a green point 
with comparable brightness to the west nearby galaxy which is rarely seen in the SDSS data.
The image size is $40\times40$~arcsec$^{2}$. 
{Alt text: Two images of the quasar field at different depths. The quasar appears fainter in the deeper image. }
}\label{fig:figinpaper_images}
\end{figure*}

\begin{figure}[!htbp]
 \begin{center}
    \includegraphics[angle=0,width=83.8mm,bb=0 0 408 408]{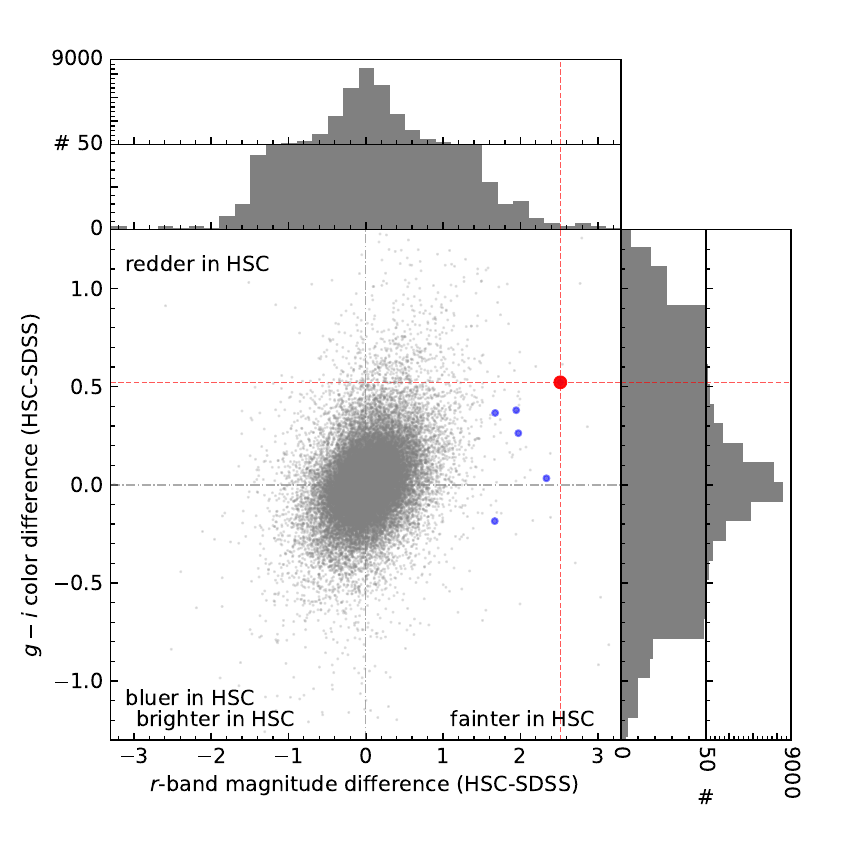}
 \end{center}
\caption{Differences of $r$-band magnitudes (in $x$ axis) and $g-i$ colors (in $y$ axis) 
of the spectroscopically identified SDSS quasars observed with HSC in the HSC-SSP program. 
The histograms for these two values are also shown at the top and right sides. 
The values for J0218-0036 are indicated as a red circle and red dashed lines. 
We also plot quasars showing $>1.5$~mag variability in the three bands 
as blue circles as reference. 
Note that several quasars showing larger variability than the quasar 
of interest in this paper are seen, especially in the histogram, and 
they do not satisfy the large variability criteria in {\it all} the three bands. 
Zero values are also shown as gray dashed lines for reference. 
An overall distribution follows the bluer-when-brighter trend 
where observed optical colors of quasars become bluer when they are brighter \citep{Paltani:1994aa,Giveon:1999aa}. 
{Alt text: 
A scatter plot of quasars showing differences in r-band magnitudes and g–i colors. Most points are densely clustered in gray. The selected quasar is highlighted, and several quasars with large variability are also highlighted. Histograms are shown along the top and right axes.
}
}\label{fig:figinpaper_magcolordiff}
\end{figure}

To search for quasars with a large long-term fading trend, 
we first compare brightness in $g$, $r$, and $i$ bands 
between the SDSS and HSC catalogs. 
The parent sample of spectroscopically identified quasars is 
the ``Final SDSS-DR14 quasar catalog''\footnote{DR14Q\_v4\_4.fits from https://www.sdss4.org/dr17/algorithms/qso\_catalog\_dr14}
\citep{paris2018}. 
The HSC magnitudes used here are the CModel magnitudes, 
{\it (gri)\_cmodel\_mag}\footnote{Coadded photometry over the survey period, 
not the measurements in the individual epochs.}, 
in the HSC Public Data Release 3 \citep{Aihara:2022vc} 
while the SDSS magnitudes are the PSF 
(point spread function) 
magnitudes in the SDSS DR14. 
The number of the SDSS quasars within the HSC footprints, 
optical variability of which are investigated in this paper, 
is \numofquasarswithinhscfootprints\ 
in the overlapped region of 421~deg$^{2}$. 
The criteria we adopt for the large variability are 
1) $>$\magvarithre~mag variability consistently in all the three filters, 
2) some extended structure in the HSC images, 
and 
3) non-detection in the 
Faint Images of the Radio Sky at Twenty-cm (FIRST; \cite{becker1995}) catalog 
obtained with 
National Radio Astronomy Observatory (NRAO) 
Very Large Array (VLA). 
The second criterion on the object extendedness is defined 
as the difference between the CModel and PSF magnitudes 
in the HSC images 
(\cite{Matsuoka:2016aa,Kawinwanichakij:2021aa}), 
and 
we adopt a loose constraint in this paper; 
$m_{\rm{CModel}}-m_{\rm{psf}}>\cmodelminuspsfthre$. 
Each of the criteria is introduced so that 
1) the effects of the inappropriate treatment of cosmic rays, bright object masks, and 
contamination of moving objects (i.e., asteroids) into the sample 
would be minimized 
2) relatively host galaxy-dominated objects 
with AGN components being faded 
would be selected 
because, in general, under the ground-based subarcsec seeing condition, 
the quasar is a point source while the host galaxy is extended, 
and 
3) effects of a relativistic jet on the observed variability would be minimized. 
Among 57 largely fading quasars satisfying the above criteria, 
we, in this paper, focus on a quasar, J0218-0036 
at (RA, Dec)$_{\rm J2000.0}$ = (\rathisobj, \decthisobj) 
at $z=$\redshiftqso\ (Figure~\ref{fig:figinpaper_images}), 
showing the largest decline among the sample; 
from 
$g_{\rm{SDSS,PSF}}=\sdssmagg\pm\sdssmagerrg$, 
$r_{\rm{SDSS,PSF}}=\sdssmagr\pm\sdssmagerrr$, and 
$i_{\rm{SDSS,PSF}}=\sdssmagi\pm\sdssmagerri$ 
in the SDSS DR14 quasar catalog  
to 
$g_{\rm{HSC,CModel}}=\hscmagg\pm\hscmagerrg$, 
$r_{\rm{HSC,CModel}}=\hscmagr\pm\hscmagerrr$, and 
$i_{\rm{HSC,CModel}}=\hscmagi\pm\hscmagerri$ 
in the HSC catalog; 
the resultant variability between the two databases are 
$\Delta{g}=\sdsstohscmagvarig\pm\sdsstohscmagvarierrorg$, 
$\Delta{r}=\sdsstohscmagvarir\pm\sdsstohscmagvarierrorr$, and 
$\Delta{i}=\sdsstohscmagvarii\pm\sdsstohscmagvarierrori$ 
(Figure~\ref{fig:figinpaper_magcolordiff}).  
The variability of this quasar is 
one of the largest amplitudes studied before. 
We confirmed that 
the small differences in the bandpasses 
between the SDSS and HSC photometric systems 
are $\lesssim1$\%\ \citep{Huang:2018aa} and confirmed not to
cause such large magnitude difference 
at $z=\zclq$ (e.g., \cite{Caplar:2020to}). 

After discovering 
the large brightness decline of the quasar, 
we quickly observed the field with 
imaging instruments 
installed in the 10.4-m 
Gran Telescopio Canarias (GTC). 
The optical instrument used is 
the Optical System for Imaging and low-Intermediate-Resolution 
Integrated Spectroscopy (OSIRIS, $r$-band) 
while the near-infrared instrument is 
Espectrógrafo Multiobjeto Infra-Rojo (EMIR; $J$ and $H$ bands; \cite{Garzon:2022aa}). 
These GTC observations motivated us 
to explore the further variability of the quasar and 
are followed by further 
observations described in \S\ref{sec:secinpaper_imgdata} 
and \S\ref{sec:secinpaper_specdata}. 

In the following subsections, 
we describe follow-up observations and archival data 
in imaging (\S\ref{sec:secinpaper_imgdata}), 
optical spectroscopy (\S\ref{sec:secinpaper_specdata}), 
and 
auxiliary data in other wavelengths (\S\ref{sec:secinpaper_anxidata}). 

\subsection{Optical and Near-Infrared Imaging Data}
\label{sec:secinpaper_imgdata}

We retrieve optical and 
near-infrared (NIR) 
photometric measurements 
from various archival catalogs. 
The brightness of J0218-0036 is also measured 
in some of the archival images and 
newly obtained optical and NIR images. 
We basically use 
``detected'' photometry while we do not use 
``undetected'' 
(``forced'') photometry 
with only exception of the Digitized Sky Survey (DSS) data. 
For the DSS, 
we basically use $5\sigma$ upper limits 
calculated by measurement errors of nearby detected sources. 

As seen in the HSC image in the right panel of Figure~\ref{fig:figinpaper_images}, 
there is a faint 
($g=22.98\pm0.01$, 
$r=22.78\pm0.01$, and 
$i=22.50\pm0.01$) 
galaxy at the west 
of J0218-0036, separated by \separcsecnearbygalaxy~arcsec. 
This galaxy is not detected in the SDSS images (i.e., not recorded in the SDSS database) 
and we basically ignore the flux contribution from this galaxy to the J0218-0036 brightness 
in shallow ($\lesssim 22$~mag) images, even with large PSFs. 
Most of the 1m-class telescope data are too shallow 
to set a meaningful constraint on the brightness of J0218-0036, 
especially in the faint phase after the 2010s. 
Therefore, we only use the detection data points in the faint phase 
measured in the images taken with larger telescopes. 

\subsubsection{Optical Imaging Data}
\label{sec:sec_optimgdata}
To construct light curves of the quasar in the optical wavelength range, 
we use photometric values 
from public catalogs and our own measurements. 
The public catalogs include 
the Palomar Transient Factory (PTF), 
the intermediate Palomar Transient Factory (iPTF), 
and 
the Zwicky Transient Facility (ZTF). 
We also measure the object brightness 
with SExtractor \citep{bertin2002} 
in the individual frames of the archival data in the field 
taken with optical wide-field imagers; 
Pan-STARRS \citep{Chambers:2016aa}, 
and 
MegaCam \citep{Boulade:2003aa} 
on the Canada-France-Hawaii Telescope (CFHT). 
We took imaging data with 
the Goodman High Throughput Spectrograph on the 4.1-m SOAR, 
OSIRIS on the 10-m GTC 
and 
LRIS on the 10-m Keck telescope. 
Imaging data with HSC 
\citep{Miyazaki:2018aa} 
are also taken in open-use filler programs 
and Planet~Nine search program 
in addition to the HSC-SSP program, 
and the object brightness is measured 
in the same way as for the archival images above. 
The aperture size is fixed to 1.0~arcsec in radius 
and flux calibration is performed 
relative to the SDSS measurements in the field. 
The light curve data in optical 
is shown in the top two panels of Figure~\ref{fig:figinpaper_lc}. 

Before the SDSS era, 
photographic plate data 
would be valuable for 
examining $\gtrsim30$-year variability of the quasar. 
Despite the relatively large measurement errors, the plate data remain valuable and warrant careful examination. 
At the position of J0218-0036, 
one measurement is recorded in the Guide Star Catalog 
(version 2.4.2; \cite{Lasker:2008aa}) 
as a source with $B_J=22.01\pm0.54$~mag at a position of (34.50793d  -0.61605d), 
with a spatial separation of $<0.1$~arcsec from the SDSS coordinate. 
The catalog is downloaded 
from the STScI DSS website, in which 3~types of magnitudes are recorded. 
We confirm that the magnitude limits are very similar to representative limits 
for the entire surveys \citep{Reid:1991ty}. 
We also conduct visual inspections of the objects in the field and 
compared the images with the measured magnitudes in the catalogs 
to verify if the limits we use in this paper are appropriate or not. 
Conversion between the plate data bandpasses and SDSS bandpasses are 
very roughly done 
from O$_{\rm{pg}}$($O$) to $g$, 
from F$_{\rm{pg}}$($R_{F}$) to $r$, 
and 
from N$_{\rm{pg}}$($I_{N}$) to $i$, 
because only rough measurements are required in this long-term brightness comparison. 
Then, 
the $5\sigma$ upper limits in the data 
are obtained to calculate the median magnitudes of objects 
with ``magerrfivesigma'' magnitude error 
within 1~arcmin around J0218-0036. 
With this DSS upper limit data, the long-term light curves 
is shown in 
Figure~\ref{fig:figinpaper_xrayplatelc}. 
Comparing the upper limits by the DSS data at MJD$\sim50,000$~days 
and the SDSS imaging fluxes, this drastic fading may have started 
sometime before the early SDSS imaging data were taken. 

\subsubsection{NIR Imaging Data} 
NIR data in the field were taken 
in the three all-sky or wide-field surveys; 
the Two Micron All Sky Survey (2MASS; \cite{skrutskie2006}), 
Large Area Survey (LAS)	of the UKIRT InfraRed Deep Sky Surveys (UKIDSS; \cite{lawrence2007}) 
with the UKIRT Wide Field Camera (WFCAM) of 
the 3.8-m United Kingdom Infra-red Telescope (UKIRT) 
and 
the VISTA Hemisphere Survey 
(VHS; \cite{mcmahon2013})
\footnote{
ESO Programme ID: 79.A-2010; 
MJD=55566.12-55566.15 corresponding to 2011-01-05; 
limiting magnitudes which are $5\sigma$ magnitude limit for point sources 
are 
$J=20.84, H=20.23, K_s=20.14$. 
}
with the NIR camera VIRCAM (VISTA InfraRed CAMera) on the 4-m 
VISTA (Visible and Infrared Survey Telescope for Astronomy; \cite{Emerson:2006aa,Dalton:2006aa}). 
No sources are detected at the quasar position 
in any of the three survey data. 

In addition to these three datasets, 
we conducted 
follow-up observations 
with 
EMIR in $J$ and $H$ bands \citep{Garzon:2022aa}
on in the 10.4-m GTC. 
The $J$ and $H$ band data were taken on 
2021-01-25 and 
2021-02-23 
with exposure times of 
1,680~sec and 1,260~sec, respectively. 
The data are reduced in a standard manner for EMIR using IRAF. 
Very marginally significant sources are found 
in either of the two filters. 
The $5\sigma$ upper limits are also obtained in the same way as done 
for the optical data as described above. 

We also took NIR imaging data with 
the Multi-Object InfraRed Camera and Spectrograph 
(MOIRCS; \cite{Suzuki:2008aa,Ichikawa:2006aa}) 
on the 8.2-m Subaru Telescope 
on 2023-08-01. 
In $Y$, $J$, and $H$ bands, 
exposures of 
$960=160\times6$~sec, 
$450=75\times6$~sec, and 
$630=35\times18$~sec 
were taken, respectively. 
The data are reduced in a standard way 
with the MCSRED2 package\footnote{https://www.naoj.org/staff/ichi/MCSRED/mcsred\_e.html}. 
The quasar is significantly detected well separated from the west nearby galaxy 
in all the three bands. 
The flux in the bands are measured relative to the PS1 ($Y$) and 2MASS ($J$ and $H$) data. 

The obtained NIR light curves are shown in the third panel of 
Figure~\ref{fig:figinpaper_lc}. 
Forced photometry at the quasar positions for the 
UKIRT, VISTA, and GTC/EMIR data 
are also plotted. 

\subsubsection{MIR Imaging Data}
\label{sec:sec_mirimg}
MIR imaging data are also available 
from 
the Spitzer Space Telescope 
and 
the Wide-field Infrared Survey Explorer (WISE; \cite{Wright:2010tg}). 

The J0218-0036 field was observed 
with the Infrared Array Camera (IRAC; \cite{Fazio:2004aa}) 
on the Spitzer Space Telescope 
in the Spitzer-IRAC Equatorial Survey (SpIES; \cite{timlin2016}) and 
the ``IRAC Imaging of Massive ACT SZ Clusters in SDSS Stripe 82'' (ACT-SZ; \cite{Menanteau2011sptz.prop80138M}) projects. 
For the SpIES, we use the coadded public catalog in the AOR-76 region 
and the recorded AB magnitudes are 
$20.07\pm0.11$~mag (AB) in the IRAC-CH1 ($3.6\mu$m) 
and 
$20.18\pm0.11$~mag (AB) in the IRAC-CH2 ($4.5\mu$m). 
The data were taken 
during a very short term, within $<1$~day, 
on the MJD of 56969~days. 
For the ACT-SZ project, the data were taken on MJD of 56200~days 
and 
no public catalogs are available. 
We conduct aperture photometry 
to measure the quasar brightness with SExtractor 
(3.4~arcsec radius aperture, 
where the PSF sizes of the images are 
$\sim1.7$~arcsec; \cite{Fazio:2004aa}) 
in the publicly available images 
so that photometry for nearby objects is consistent with the SpIES coadded catalog. 

Since the start of the WISE mission, 
J0218-0036 field was observed in the 
ALLWISE\footnote{
ALLWISE is a combined infrared data release from the original WISE mission, 
incorporating observations from both the cryogenic and post-cryogenic phases (2010–2011), 
but excluding data from the later NEOWISE reactivation phase.
} 
\citep{Cutri:2013aa} 
and 
Near-Earth Object WISE Reactivation (NEOWISE-R; \cite{Mainzer:2011vk,Mainzer:2014vt})
projects. 
In the preliminary version \citep{Eisenhardt:2020wo} of the 
CATWISE2020 coadded catalog \citep{Marocco:2021uk}, 
where all the WISE data are coadded, 
there is a measurement record 
at the position (RA, Dec)=($34.50778$, $-0.61598$) [deg] 
\footnote{
The positional error is $0.26$~arcsec, 
which is much less than the PSF sizes in the WISE bands. 
}
of 
W1=$17.68\pm0.10$~mag (3.4$\mu$m) 
W2=$16.74\pm0.13$~mag (4.6$\mu$m) 
at a mean observation MJD of $56670.965492$~days. 
The observing baseline is as long as $\sim10$~years, 
so it is useful to examine its time variability 
with more time-divided data. 
However, 
the single-epoch data easily available on the IRSA archive 
are too shallow and 
the detection is not as significant as $<3\sigma$ 
even in the most significant cases. 
Then, we utilize the WISE/NEOWISE Coadder interface to create multi-epoch W1 and W2 images 
of $18\times18$~arcmin$^{2}$ 
with 2-year observing baseline 
from 2010-01-01 to 2022-12-31 
(except for the years 2012 and 2013), 
resulting in 6~epochs. 
We conduct aperture photometry of all the detected objects in the images 
with SExtractor. 
Flux calibration is done using the magnitudes 
in the ALLWISE catalog as a reference. 
The angular resolutions of the W1 and W2 bands are 6.1~arcsec and 6.4~arcsec, respectively. 
The aperture size is set to be 9.6~arcsec in diameter in both the bands. 
The nearby west 
galaxy is inevitably included in the aperture for the quasar, 
but the contamination from the nearby galaxy is ignored. 
Note that the detected position is more consistent with the quasar coordinate 
rather than the nearby galaxy coordinate. 

No significant sources are recorded 
in the MIPS of the Spitzer Space Telescope 
or 
in the W3 or W4 band of the WISE. 

\begin{figure*}
 \begin{center}
    \includegraphics[width=165.5mm,bb=0 0 503 503]{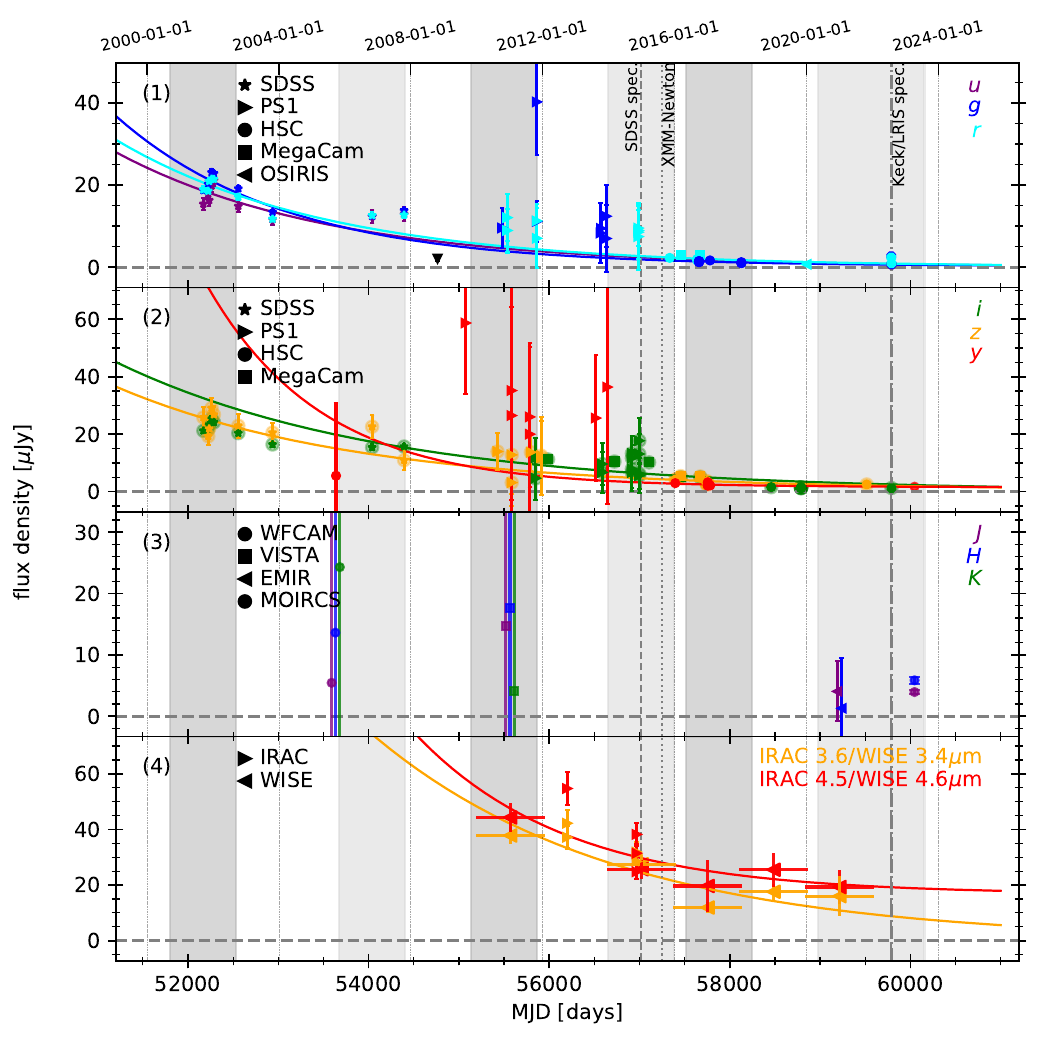}
 \end{center}
\caption{
Multi-band light curves from the SDSS imaging epochs to the present. 
From top to bottom, 
(1) $u$, $g$, and $r$-band data, 
(2) $i$, $z$, and $y$-band data, 
(3) $J$, $H$, and $K_s$-band data, 
and 
(4) 3.6/3.4$\mu$m and 4.5/4.6$\mu$m (IRAC/WISE) data 
are shown in the first, second, third, and fourth panels, respectively. 
Symbols and colors indicate 
instruments and filters, respectively. 
Note that, in the NIR (third) panel, only the rightmost data points are with significant detection. 
The three vertical lines indicate 
the epochs of the SDSS spectroscopy (dashed), XMM-Newton observations (dotted), 
and Keck/LRIS spectroscopy (long-dashed), from left to right. 
Six shaded regions indicate 
the periods for multi-epoch SED fitting 
done in \S\ref{sec:secinpaper_seddecompose} (Figure~\ref{fig:figinpaper_bolLEddRatio}). 
We fit these photometry values 
using an exponential decay 
($f=f_{0}\exp[-(t-t_{0})/\tau$]) for each band. 
{Alt text: 
Four-panel light curves showing flux density over time across multiple optical, 
NIR, and MIR bands. 
Fitted exponential decay trends are overlaid on the data. 
Six periods of interest for the SED fitting and epochs of 
spectroscopic observations are indicated 
as gray shaded regions. 
}
}\label{fig:figinpaper_lc}
\end{figure*}

The obtained MIR light curves are shown in the fourth panel of Figure~\ref{fig:figinpaper_lc}. 
According to the comparison between WISE and Spitzer/IRAC magnitudes 
done in \citet{Jarrett:2011aa}, the 
bandpass differences between Spitzer/IRAC and WISE 
only cause 2-3\% level difference in magnitude and can be ignored in this paper.

\begin{figure*}
 \begin{center}
    \includegraphics[width=135mm,bb=0 0 531 358]{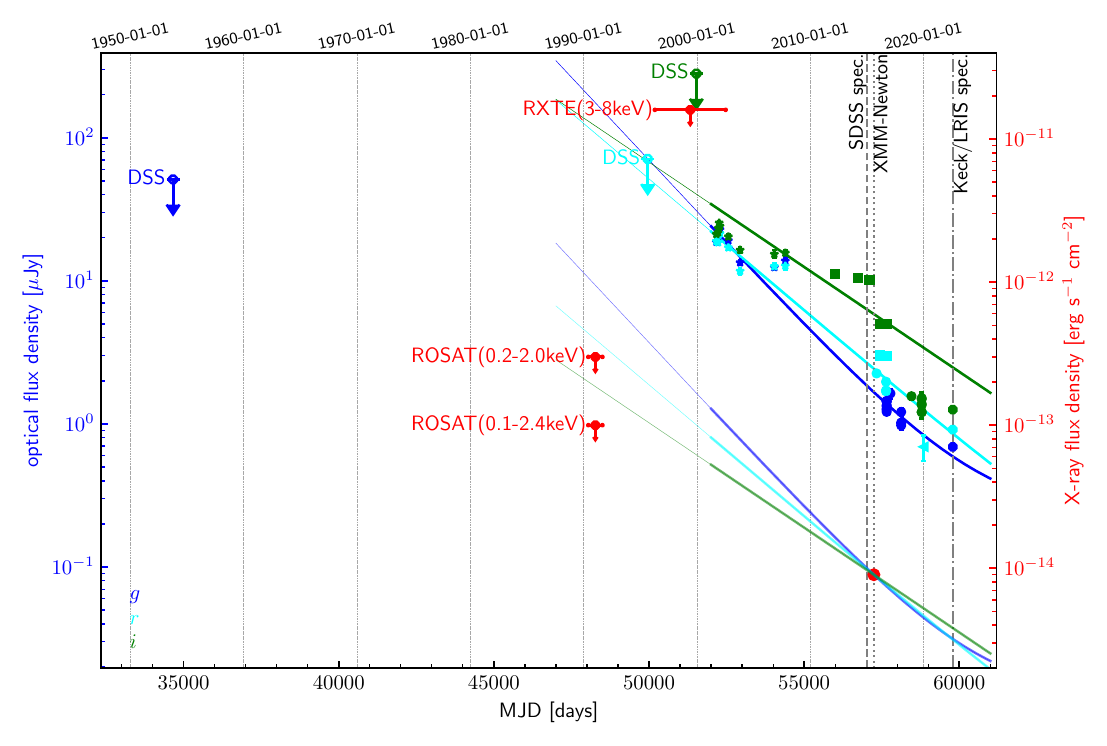}
 \end{center}
\caption{
Long-term multi-band light curves from the 1950s to the present 
in short optical ($g$, $r$, and $i$ bands in blue, cyan and green, respectively) 
for the left-hand axis 
and X-ray in red for the right-hand axis. 
The above fitted lines for the three-band optical data are the same as 
those (exponential decay) in Figure~\ref{fig:figinpaper_lc}. 
The bottom three lines are scaled lines of the above lines 
to match the XMM-Newton X-ray measurement in 2015. 
Thin lines indicated extrapolated ranges without any data for the fitting. 
Before 2000, we also show $3\sigma$ upper limits 
in optical from the plate data 
(blue, cyan, and green) 
and 
in X-ray from ROSAT and RXTE (red). 
{Alt text: 
Long-term light curves showing optical and X-ray flux densities over time from the 1950s to the present. Fitted exponential decay trends for the optical data are overlaid. Upper limits are indicated for early epochs. Epochs of spectroscopic observations are marked.
}
}\label{fig:figinpaper_xrayplatelc}
\end{figure*}

In summary, the series of the MIR imaging data indicate 
that the quasar is declining its brightness 
over the 
13~years in the observed frame.

\subsection{Optical Spectroscopic Data}\label{sec:secinpaper_specdata}
For the quasar, an optical spectrum was taken in the SDSS project 
as described in \S\ref{sec:secinpaper_sdssspecdata}. 
In addition, we took two optical spectra of this quasar in 2022 
as described in \S\ref{sec:secinpaper_ourownspecdata}. 

\subsubsection{SDSS Spectrum}\label{sec:secinpaper_sdssspecdata}

The first-epoch spectrum of the quasar 
was taken in the Extended Baryon Oscillation Spectroscopic Survey 
(eBOSS; \cite{Dawson:2016aa}) project 
at MJD=\mjdsdssspec\ 
during the rapidly fading phase 
as indicated in the dashed line in Figure~\ref{fig:figinpaper_lc}. 
Broad C\,{\sc iv}, C\,{\sc iii}], and Mg\,{\sc ii} emission lines are detected and 
the redshift of $z=\zclq$ was determined 
with a $\chi^{2}$ fitting for the template \citep{Stoughton:2002aa}. 
Flux calibration for the SDSS spectra is done 
with simultaneously observed SDSS standard stars 
and to match the colors measured in the SDSS images. 
We confirm that the flux calibration of the spectrum 
is consistent with the photometry 
for further discussion in later sections 
by comparing the absolute flux of the spectrum 
with the expected (interpolated) brightness from the long-term light curves 
shown in Figure~\ref{fig:figinpaper_lc}. 

A single-epoch black hole mass \bhmass\ 
is estimated in two papers; 
\citet{rakshit2020} for the DR14 spectrum 
and 
\citet{Wu:2022aa} for the DR16 spectrum. 
They use 
continuum luminosity of the calibrated spectra and 
Mg\,{\sc ii} \citep{wang2009_mg2bhmass,Vestergaard:2009aa,shen2011} and 
C\,{\sc iv} \citep{Vestergaard:2006aa} 
emission lines to calculate 
\bhmass\ of the quasar 
and obtain consistent values 
as summarized below; 
$\log$\bhmass$=$ 
\logbhmassmgtworakshit\ (DR14, Mg\,{\sc ii}), 
\logbhmassmgtwowu\ (DR16, Mg\,{\sc ii}), 
\logbhmasscfourrakshit\ (DR14, C\,{\sc iv}), 
and 
\logbhmasscfourwu\ (DR16, C\,{\sc iv}). 
Note that different calibration parameters are used, 
\citet{Vestergaard:2006aa} in \citet{rakshit2020} and \citet{shen2011} in \citet{Wu:2022aa} for C\,{\sc iv} 
while the same parameters are used for Mg\,{\sc ii} \citep{Vestergaard:2009aa}. 

These estimates of the BH mass could be a lower limit. 
From the exponential decay fitting for the light curves, 
the decline factor between the earliest SDSS imaging epochs and the SDSS spectroscopic epoch is 
$\sim\declinefactorsdssimgtospecnew$ 
(Figure~\ref{fig:figinpaper_lc}). 
The virial theorem simply expects line widths of broad lines scale to the continuum luminosity 
with a power-law index of $-1/4$ (FWHM$\propto L^{-1/4}$), 
but it is not always the case. 
Contrary to broad H$\beta$ emission lines following 
a linewidth-continuum lumionisty relation expected from the virial theorem 
(``breathing'', \cite{Wang:2020aa}), 
broad C\,{\sc iv} or Mg\,{\sc ii} emission lines do not follow such a simple relation. 
Then, the derived BH masses for the quasar 
estimated in \citet{rakshit2020} and \citet{Wu:2022aa} 
could be highly uncertain. 
As a reference, if BH mass 
can be naively scaled with the square root of the continuum luminosity, 
the BH mass would be $\sim\correctedforfading$ times 
as large as values in the previous papers, 
$\log$\bhmass$=$
\bhmassfiducialcorrectedforfading\ 
if a spectrum was taken around the early SDSS imaging epochs. 

In the \citet{rakshit2020} and \citet{Wu:2022aa} papers, 
many other parameters are measured in the SDSS/eBOSS spectrum. 
The bolometric luminosity $L_{\rm{bol}}$ and 
Eddington ratio 
$\lambda_{\rm{Edd}}$ 
are also calculated to be 
$10^{45.2}$ and $10^{45.3}$ [erg~s$^{-1}$], 
and $10^{-1.17}$ and $10^{-1.25}$, respectively. 
The bolometric correction of 5.15 
for 3000\AA\ luminosity \citep{shen2011,Richards:2006aa} 
is assumed in these calculations. 
Other measurements for the broad lines such as 
full widths at half maximum (FWHMs) 
and line fluxes are 
summarized in Table~\ref{tab:tabinpaper_specprop} 
and all of these are consistent within the errors 
between the two measurements. 
All these values 
will be compared to those measured 
in our Keck/LRIS spectrum (\S\ref{sec:secinpaper_ourownspecdata}).

\subsubsection{Additional Spectra}\label{sec:secinpaper_ourownspecdata}

\begin{figure*}
 \begin{center}
    \includegraphics[width=187.95mm,bb=0 0 701 408]{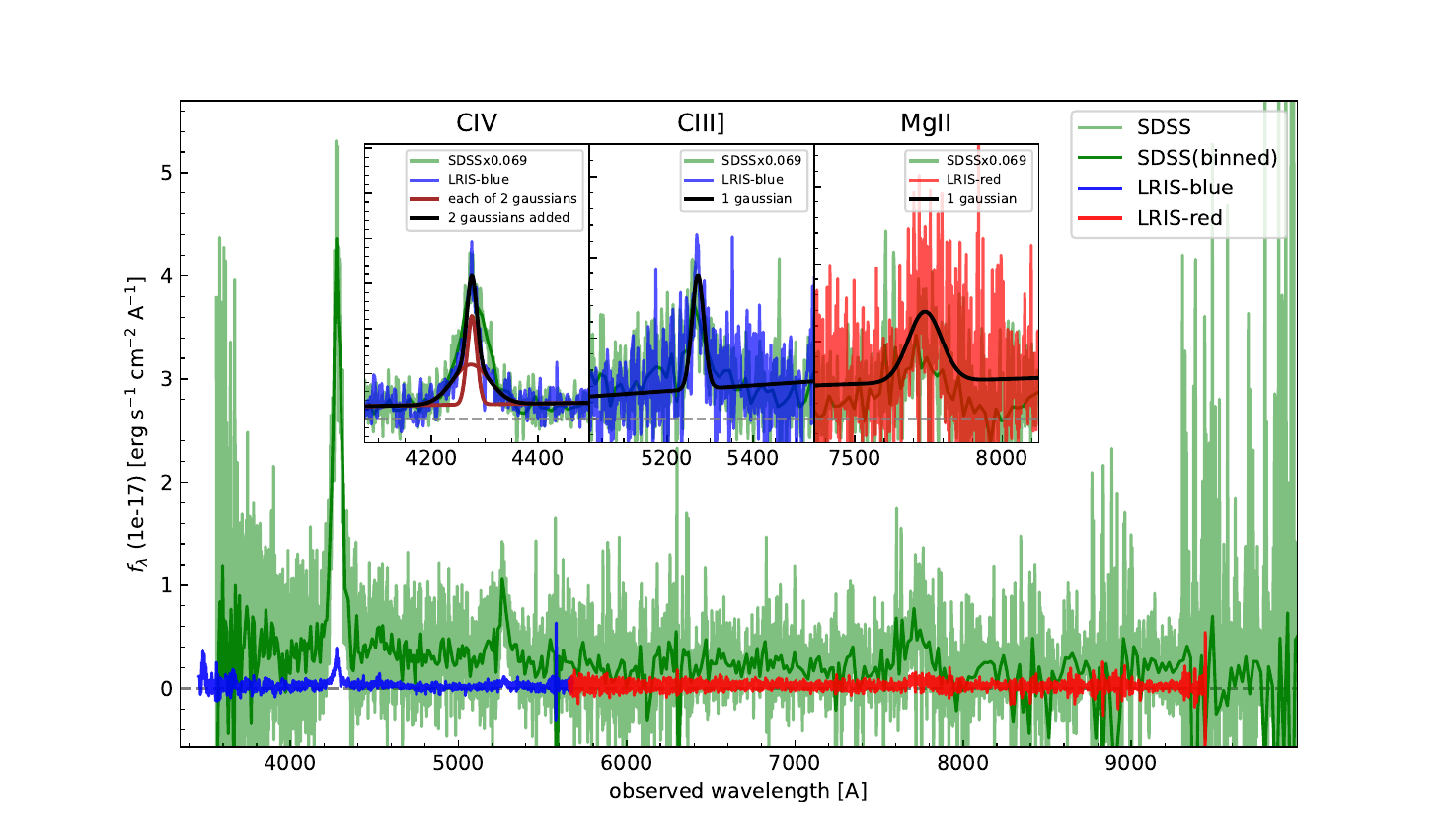}
 \end{center}
\caption{
Spectra of J0218-0036 in 
green (SDSS), 
blue (Keck/LRIS, blue arm), 
and 
red (Keck/LRIS, red arm). 
The binned SDSS spectrum is also shown in green. 
The three insets are magnified views of the C\,{\sc iv}, C\,{\sc iii}], and Mg\,{\sc ii} emission lines 
with relative scaling of the SDSS specrum to the LRIS spectrum 
by a factor of $\sim16$. 
For the C\,{\sc iv} emission line in the LRIS spectrum, 
two fitted gaussians are shown in brown 
and their sum is shown in black 
while for the C\,{\sc iii}] and Mg\,{\sc ii} emission lines, the fitted one gaussian is shown in black. 
{Alt text: 
Optical spectra showing flux density versus observed wavelength for the source J0218-0036. Spectra from SDSS and Keck/LRIS are displayed, with an inset magnifying 
the C\,{\sc iv}, C\,{\sc iii}], and Mg\,{\sc ii} 
emission lines. Gaussian fits to the emission lines are also shown in the insets.
}
}\label{fig:figinpaper_lrissdssspeccomp}
\end{figure*}

We obtained optical long-slit spectra of 
the quasar and the nearby galaxy 
with 
Low Resolution Imaging Spectrometer (LRIS; \cite{oke1995,Rockosi:2010tv}) 
on the Keck-I telescope 
over 2~nights, 
on July 31 and August 1, 2022. 
The dichroic mirror D560 was used. 
The grism 400/3400 and no filter were used in the blue side 
and 
the grating 400/8500 and the GG495 filter were used in the red side. 
Then, the covered wavelength range is 
3500-9400~\AA. 
The atmospheric dispersion corrector was used. 
The spectral and spatial binnings are 
$2\times2$ and $2\times1$ in the blue and red sides, respectively. 
The total exposure times were 9900~sec ($11\times900$-sec exposures). 
The slit width is 
1.0~arcsec 
and the resultant spectral resolution is $R=600-1000$. 
A spectrophotometric standard star used for the flux calibration 
is Hz4. 
The LRIS data is reduced with {\ttfamily PypeIt},\footnote{\url{https://pypeit.readthedocs.io/en/latest/}}
a Python package for semi-automated reduction of 
astronomical slit-based spectroscopy
\citep{Prochaska:2020ts,Prochaska:2020wq}. 
The observed Keck/LRIS spectrum is shown 
in Figure~\ref{fig:figinpaper_lrissdssspeccomp} 
and discussed in later sections. 

\begin{table*}[!htbp]
  \tbl{Measured properties of the SDSS \citep{rakshit2020,Wu:2022aa} and Keck/LRIS spectra. ``BR'' indicates the measurements for the broad emission lines. 
  The contribution from the host galaxy component is ignored to calculate these values. 
  The units of FWHMs, line fluxes, black hole masses, and Eddington ratios are \kmpersec, \fluxunit, $M_{\odot}$, dimensionless, respectively. 
  These two spectra were taken at epochs separated by $\sim2.7$~years in the rest-frame ($\sim7.6$~years in the observed frame). 
  }{
  \begin{tabular}{llllllllll}
      \hline
      paper & \citet{rakshit2020}   & \citet{Wu:2022aa}     & this work\\\hline
      data  & SDSS/DR14 & SDSS/DR16 & Keck/LRIS\\\hline
      MJD         & \mjdsdssspec & \mjdsdssspec & \mjdlrisspec \\
      \hline
      FWHM(C\,{\sc iv})   & $3940.50\pm287.40$ & $4557.79\pm547.68$ & $2542.17\pm98.93$ (all), $5739.19\pm3921.97$ (BR)\\
      FWHM(C\,{\sc iii}]) & $2106.69\pm708.45$ & $2072.62\pm685.88$ (all), $2121.21\pm472.81$ (BR) & $1844.41\pm642.22$\\
      FWHM(Mg\,{\sc ii})  & $3692.52\pm1782.45$ & $3171.51\pm1198.94$ (all), $4620.67\pm1118.85$ (BR) & $5171.09\pm4906.01$\\
      \hline
      flux(C\,{\sc iv})   & $3.66^{+0.30}_{-0.28}\times10^{-15}$ & $(2.90\pm0.07)\times10^{-15}$ & $(9.6\pm0.96)\times10^{-17}$\\
      flux(C\,{\sc iii}]) & $8.75^{+1.41}_{-1.21}\times10^{-16}$ & $(5.08\pm0.51)\times10^{-16}$ (all), $(5.08\pm0.71)\times10^{-16}$ (BR) & $(2.4\pm0.24)\times10^{-17}$\\
      flux(Mg\,{\sc ii})  & $7.64\pm1.28\times10^{-16}$ & $(7.43\pm1.18)\times10^{-16}$ & $(6.4\pm1.6)\times10^{-17}$\\
      \hline
      $\log M_{\rm{BH,CIV}}$  & $8.25\pm0.42$ & $8.23\pm0.11$ & $8.26\pm0.03$, $7.64\pm0.11$\\
      $\log M_{\rm{BH,MgII}}$ & $8.09\pm0.07$ (fiducial) & $8.44\pm0.21$ (fiducial) & $9.08\pm0.82$\\
      \hline
      $\log L_{\rm{bol}}$     & $45.22$ & $45.30$ & $44.10$\\
      \hline
      $\log \lambda_{\rm{Edd}}$ & $-1.17$ & $-1.25$ &  $-2.40$\\
      \hline
    \end{tabular}}\label{tab:tabinpaper_specprop}
\begin{tabnote}
\end{tabnote}
\end{table*}

An additional 5700-s spectrum ($1,200\times4+900\times1$) 
was obtained with 
the Goodman Spectrograph on the 4.1-m SOAR telescope 
on July 28, 2022 (UT). 
The multi-object spectroscopy mode was applied 
with the slit widths of 1.0~arcsec. 
The grating 400M1 (400~lines~mm$^{-1}$) was used with no order-sort filter, 
giving the wavelength range 
from 3200\AA\ to 6400\AA, 
to cover the C\,{\sc iv} and C\,{\sc iii}] lines of the quasar, 
with a spectral resolution $R$ of $\sim1000$. 
The atmospheric dispersion corrector was used. 
The obtained data are too shallow due to the bad seeing condition during the exposures 
and no significant emission (continuum or emission lines) is 
detected at the quasar position. 

Figure~\ref{fig:figinpaper_lrissdssspeccomp} shows 
a comparison of 
the SDSS/eBOSS spectrum and 
the Keck/LRIS spectrum 
taken at epochs separated by 
$\sim2.7$~years in the rest-frame 
($\sim7.6$~years in the observed frame). 
As easily seen in the figure, 
C\,{\sc iv}, C\,{\sc iii}], and Mg\,{\sc ii} broad lines in the LRIS spectrum 
are still as broad as those 
in the SDSS spectrum 
while the continuum significantly reduce its brightness. 
For the strong broad emission lines 
of C\,{\sc iv}, C\,{\sc iii}], and Mg\,{\sc ii}, 
the LRIS spectrum $f_{\lambda}(\lambda)_{\rm{LRIS}}$ 
is fitted with a combination of 
gaussian function and a linear polynomial 
around the lines of interest with $\pm100$~\AA\ range. 
For C\,{\sc iv}, two gaussians are used for the fitting 
while the previous fitting results for the SDSS spectrum 
\citep{rakshit2020,Wu:2022aa} 
uses three gaussians 
simply because of the limited signal-to-noise ratios of the LRIS spectrum 
in the much fainter phase. 
The measured properties for the two spectra are summarized in Table~\ref{tab:tabinpaper_specprop}. 
Fluxes of the C\,{\sc iv} and C\,{\sc iii}] emission lines in the Keck/LRIS spectrum 
are consistently weaker than those in the SDSS spectrum 
by factors of $\sim30-38$ and $\sim21-36$, 
respectively. 
The Mg\,{\sc ii} line and the continuum are also fainter 
by factors of $\sim11-12$ and $\sim16$, respectively, 
although the Mg\,{\sc ii} line in the LRIS spectrum is noisy. 
The smaller continuum decline relative to the broad-line 
declines can be partly attributed to reverberation 
time delays (a few tens of days in the rest frame; \cite{Grier:2019aa}), 
and the Baldwin effect may also contribute \citep{Ross:2020aa}. 
Given the short expected delay for C\,{\sc iv} 
($\lesssim10$~days; \cite{Kaspi:2007aa,Lira:2018aa}), 
time-delay effects are likely minor for that line. 

The widths of the lines are not changed significantly, or 
even show anti-breathing trend, 
which is consistent with results obtained 
in the SDSS Reverberation Mapping (RM) project 
\citep{Wang:2020aa} 
although there is a large difference in variability amplitude 
between SDSS-RM quasars and the quasar in this paper.  

The BH mass is also estimated from the LRIS spectrum with the single-epoch spectrum method, 
independently from the SDSS spectrum 
done in \citet{rakshit2020} and \citet{Wu:2022aa} (see Section~\ref{sec:secinpaper_sdssspecdata}). 
The equation to calculate the BH mass is the same as that in \citet{rakshit2020}, 
\begin{eqnarray}
    \log\left(\frac{M_{\rm{BH}}}{M_{\odot}}\right)=A+B\log\left(\frac{\lambda L_{\lambda}}{10^{44} \rm{erg~s}^{-1}}\right)+2\log\left(\frac{\rm{FWHM}}{\rm{km s}^{-1}}\right)
\end{eqnarray}
where $(A,B)=(0.860, 0.50)$ for CIV \citep{Vestergaard:2006aa} 
and $(A,B)=(0.860, 0.50)$ for MgII \citep{Vestergaard:2009aa}, respectively. 
The obtained BH masses with the LRIS spectrum are 
$\log(M_{\rm{BH}})=8.26\pm0.03, 7.64\pm0.11$ for C\,{\sc iv} 
and 
$\log(M_{\rm{BH}})=9.08\pm0.82$ for Mg\,{\sc ii}. 
Given that the BH masses estimated with the spectra at the two epochs 
is consistent with each other, 
we adopt the BH mass of the quasar to be $\log(M_{\rm{BH}})=8.2$ in later discussions. 
The bolometric luminosity $L_{\rm{bol}}$, derived by multiplying 
the rest-frame 3000~\AA\ luminosity by a factor of $5.15$ \citep{Richards:2006aa} as done in the two previous papers, 
and the Eddington ratio $\lambda_{\rm{Edd}}$ 
at the LRIS spectroscopy epoch 
are calculated to be 
$\log L_{\rm{bol}}=44.10$, 
and 
$\log \lambda_{\rm{Edd}}=-2.40$, 
respectively. 

The equivalent width (EW) of the C\,{\sc iv} emission line in the SDSS spectrum 
is as large as $\sim400$~\AA\ in the rest-frame \citep{rakshit2020} 
and almost at the larger edge of the C\,{\sc iv} EW distribution of the quasars. 
The C\,{\sc iv} EW of the SDSS quasar composite spectrum 
is also as small as $23.78\pm0.10$~\AA\ \citep{vandenberk2001}. 
In the LRIS spectrum, the C\,{\sc iv} EW decreases to $\sim92$~\AA. 
This decrease in EW shows a trend opposite to the Baldwin effect. 
This may indicate the time delay of the C\,{\sc iv} broad line emission line flux 
respect to the continuum. 

The line flux ratio of C\,{\sc iii}]/C\,{\sc iv} is also worth being investigated 
especially compared to those of quasars 
because the LRIS spectrum of this quasar 
shows 
the broad C\,{\sc iv} 
and 
C\,{\sc iii}] lines. 
This ratio is sensitive to ionization parameters \citep{Matsuoka:2009aa,Nagao:2006ac}. 
The C\,{\sc iii}]/C\,{\sc iv} flux ratio of this quasar is $0.18-0.24$ in the SDSS spectrum 
and $0.25$ in the LRIS spectrum. 
These ratios are in a small part of the entire distribution 
using the \citet{rakshit2020} 
measurement ($0.5-1.0$) 
and 
smaller than those measured 
in the redshift- and luminosity-binned composite spectra of SDSS quasars 
(0.3–0.4; \cite{Nagao:2006ac}), 
indicating a large ionization parameter. 
At the two spectroscopic epochs, 
the rest-frame UV brightness has significantly 
faded (Figure~\ref{fig:figinpaper_lc}), however, 
this quasar sustains a high ionization parameter, 
possibly indicating the existence of 
a lower density gas at a closer distance to the central SMBH 
than typical Type-1 AGNs. 

\subsection{Auxiliary Data}\label{sec:secinpaper_anxidata}

\subsubsection{X-ray}
\label{sec:secinpaper_xraydata}

This object is recorded in the XMM-Newton archival catalog. 
XMM-Newton data are searched 
in the 4XMM-DR12 catalog\footnote{http://xmm-catalog.irap.omp.eu/} and 
one detection, 4XMM~J021801.8-003658, is recorded\footnote{http://xmm-catalog.irap.omp.eu/source/207622902010022}\footnote{http://xmm-catalog.irap.omp.eu/detection/107622902010022} around the quasar coordinate. 
The MJD of the data ranges 
from 57247.036 to 57247.306 (roughly on 2015-08-13). 
The measured fluxes are 
$(3.08\pm0.72)\times10^{-14}$ [\fluxunit] in the 0.2-12~keV energy range 
and 
$(1.34\pm0.14)\times10^{-14}$ [\fluxunit] in the 0.5-4.5~keV energy range. 
We fit the XMM-Newton spectrum 
with a single power-law 
with 
the web tool {\it WebSpec}\footnote{https://heasarc.gsfc.nasa.gov/webspec/webspec.html} 
and obtain 
an intrinsic neutral hydrogen column density 
$N_H$ of 
$2.6\times10^{21}$ [cm$^{-2}$], 
indicating that the quasar is not obscured 
in the XMM-Newton observation epoch in 2015. 
The Galactic $N_H$ towards the quasar 
of $\sim2.78\times10^{20}$ [cm$^{-2}$] 
is much smaller than the intrinsic $N_H$ measured above 
and we ignore the Galactic $N_H$ in further discussion. 
In the calculation of {\it WebSpec}, we adopt
the composite X-ray spectrum of type-1 AGNs (Figure~2 in \cite{Kawaguchi:2001aa}). 

Earlier X-ray activity, before the XMM observation above, 
can be investigated as below. 
Almost the entire sky was observed with 
the ROentgen SATellite (ROSAT) and 
the Rossi X-ray Timing Explorer (RXTE) satellite. 
The source catalogs are made publicly available; 
ROSAT All-Sky Catalog (2RXS; \cite{Boller:2016uo,Voges:2000wn}) and
RXTE All-Sky Slew Survey Catalog \citep{Revnivtsev:2004aa}. 
The quasar is not detected in either of the satellites 
and the obtained upper limits are 
$1\times10^{-13}$ 
[\fluxunit] in the 0.1-2.4~keV for ROSAT (2RXS) and
$1\times10^{-11}$ 
[\fluxunit] in the 3-8~keV band for RXTE.
When we assume the same intrinsic $N_H$ of 
$2.6\times10^{21}$ [cm$^{-2}$] 
as measured in the XMM-Newton data, 
the RXTE's upper limit is converted to be 
$2.6\times10^{-11}$ [\fluxunit] 
in the 0.12-2.11~keV 
using the web tool 
{\it WebSpec} 
In other words, 
the ROSAT limit is tighter than RXTE by a factor of $\sim260$. 

All these three flux constraints
(the detection with XMM-Newton and the upper limits of ROSAT and RXTE) 
are plotted in Figure~\ref{fig:figinpaper_xrayplatelc}. 
For the XMM-Newton measurement, we sum 
the three energy band fluxes (0.2-0.5~keV, 0.5-1.0~keV, and 1.0-2.0~keV) and 
obtain $9.0\times10^{-15}$ [\fluxunit]
in the 0.2-2.0~keV. 
As a reference, we also plot 
optical ($g,r,i$-band) data with 
the fitted curves shown in Figure~\ref{fig:figinpaper_lc}. 
If we normalize the fitted curve to the XMM-Newton flux 
at the XMM-Newton observing date and assume the same variability behavior  
between the rest-frame UV and X-ray, the ROSAT and RXTE upper limits 
can be compared to X-ray flux expected by the fitted curve. 
The non-detection in the ROSAT and RXTE data 
is consistent with the fitted curve, 
implying that the quasar was not brighter  
than the SDSS imaging epochs 
(MJD$~\sim$52000-54000~days) by more than 
a factor of $\sim10$ 
in the ROSAT and RXTE observation epochs. 
These upper limits 
are also consistent 
with the extrapolated light-curve and 
the inferred onset of the fading discussed in \S\ref{sec:sec_optimgdata}.

We also search for detections in the archives of 
Chandra Source Catalog Release 2.0 (CSC 2.0; \cite{Evans:2024aa}), 
NuSTAR Serendipitous Survey catalog \citep{Greenwell:2024aa}, 
The 7-year MAXI/GSC X-Ray Source Catalog (3MAXI; \cite{Kawamuro:2018aa}), 
and 
the X-ray source catalog detected with 
the Mikhail Pavlinsky ART-XC telescope on board the SRG observatory 
during the first operation year 
\citep{Pavlinsky:2022aa}, 
but no X-ray detections are recorded in any of the above catalogs.

\subsubsection{radio}
J0218-0036 is not detected in any public radio catalogs, 
including 
the FIRST (1.4~GHz, rms of 0.130~mJy; \cite{becker1995}), 
the NRAO VLA Sky Survey (NVSS, 1.4~GHz, \cite{condon1998AJ....115.1693C}), 
the VLA Low-Frequency Sky Survey Redux (VLSSr; 74~MHz, 
rms of 0.1~Jy~beam$^{-1}$, 
\cite{Lane:2014aa,Cohen:2007th}),  
VLA Sky Survey (VLASS; \cite{lacy2020})\footnote{https://science.nrao.edu/vlass}, 
and 
Rapid ASKAP Continuum Survey (RACS; \cite{McConnell:2020aa})
\footnote{
RACS-low in 887.5~MHz \citep{Hale:2021aa} and 
RACS-mid in 1367.5~MHz \citep{Duchesne:2023aa}.
}. 

We also conducted a follow-up radio observation 
to explore a possibility of a new jet launch 
because more jet powers are observed 
for AGNs with lower Eddington ratios in general \citep{Rusinek:2017aa}. 
We carried out the 8.4~GHz observation 
on Jan. 20, 2023 
with the Yamaguchi Interferometer (YI) 
consisting of two radio telescopes with diameters of 32~m and 34~m 
separated by the baseline length of 108.7~m \citep{Fujisawa:2022aa}. 
We also observed both radio sources 3C~48 and J0216-0105 
as a flux calibrator and a gain calibrator, respectively. 
The integration time of the quasar is 40~min 
and it is not detected with a $3\sigma$ upper limit of 2.6~mJy.

Radio-loudness $R_i\equiv\log_{10}(f_{\rm{1.4GHz}}/f_{i})$ defined 
at the observed frame 
is a ratio of the 1.4~GHz flux 
to the $i$-band flux 
without K~corrections \citep{ivezic2002}. 
According to the FIRST Catalog Search 
website\footnote{http://sundog.stsci.edu/cgi-bin/searchfirst}, 
we find that 
the catalog detection limit (including CLEAN bias) at the position is 0.88~mJy/beam. 
The typical FIRST rms noise is 0.130~mJy, corresponding to $3\sigma$ limit of $0.390$~mJy. 
These two are different from each other only by a factor of $\sim2.3$, and 
the obtained $R_i\equiv\log_{10}(f_{\rm{1.4GHz}}/f_{i})$ is 0.0-0.2 
if we use $i$-band magnitude of 20.8 
measured in the SDSS imaging epochs. 
Note that we here ignore the time difference 
of $\sim10$ years between the SDSS and VLA observations. 
In the faint phase, the radio loudness calculated from the HSC $i$-band magnitude 
and the YI radio $3\sigma$ limit is 
$R_i\equiv\log_{10}(f_{\rm{1.4GHz}}/f_{i})$ of 0.84. 
All these radio-loudness values are well below those 
for the FIRST-SDSS \citep{ivezic2002} 
and FIRST-HSC sources \citep{yamashita2018} 
by more than 1 order of magnitude. 
In summary, there is no strong evidence that 
the quasar originally belongs to a radio-loud population. 
There is also no evidence that 
this fading phenomenon is accompanied 
with a new radio activity, for example, 
a new jet ejection as radio activity increases 
observed for state-changing stellar-mass 
black holes \citep{Fender:2009aa,Belloni:2010aa}. 

\section{SED Decomposition and Origin of the Large Flux Decline}\label{sec:secinpaper_seddecompose}

We construct a temporal series of the SEDs 
from optical to MIR in the observed frame 
of this quasar to see how the SEDs change over time. 
We plot photometry measurements 
taken in six 2-year windows around 
(1) the early phase of the SDSS imaging epochs, 
(2) the late phase of the SDSS imaging epochs, 
(3) the early WISE imaging epochs, 
(4) the SDSS spectroscopy epoch, 
(5) the HSC-SSP imaging epochs, 
and 
(6) the Keck/LRIS spectroscopy epoch. 
As seen in Figure~\ref{fig:figinpaper_lc}, 
the rapid large fading already started around the epochs (1) and (2). 

We decompose each SED 
into two components; an AGN and a host galaxy. 
The former, AGN, component can be variable in time 
while the latter, galaxy, component should be constant in time. 
We consider two 
scenarios explaining the variability of the AGN components: 
(i) large intrinsic luminosity variability of the AGN component itself 
(i.e., emission from the accretion disk and dusty torus)
and 
(ii) small normal ($\sim0.2$~mag; \cite{vandenberk2004}) 
intrinsic luminosity variability with a large change in obscuring dust 
in its line-of-sight, 
causing a large temporal change in dust extinction for the AGN component. 
In other words, 
the AGN component color does not change over time in the scenario (i) 
while it changes much in the scenario (ii) because of the large dependence of 
dust extinction on wavelength. 
We here ignore the bluer-when-brighter trend 
usually observed in observed optical wavelengths 
which shows much smaller color changes than those 
by large variability in dust extinction. 
We also ignore a possible state change from standard disk SED to LINER-like SED for simplicity. 
We will discuss this later when we estimate the time variability of the Eddington ratio. 

The AGN spectral template used for the SED fitting 
is that of \citet{Bianchini:2019wu} 
to cover the wide wavelength range from UV to NIR in the rest-frame. 
We note that the spectral template of \citet{vandenberk2001} does not 
cover the entire wavelength region 
in our data. 
Extinction law for the AGN component used in the scenario (ii) 
is that for 
Small Magellanic Cloud (SMC; \cite{Gordon:2003aa}) 
rather than that of the Milky Way with the $\sim2100$\AA\ bump 
as usually used for extinction studies in AGNs \citep{Richards:2003aa}. 
The range of $A_V$ is $0-10$~mag. 

The galaxy spctral templates are taken from 
the Flexible Stellar Population Synthesis 
(FSPS; \cite{Conroy:2009aa,Conroy:2010ac}). 
The stellar isochrone of 
Mesa Isochrones and Stellar Tracks (MIST; \cite{Dotter:2016aa,Choi:2016aa}) 
is adoped. 
The initial mass function used in the FSPS is 
from \citet{Chabrier:2003aa}. 
Star formation history (SFH) adopted is $\tau$-model 
[SFR$(t)$$\propto\exp(-t/\tau)$, with a fixed $\tau$ of 1~Gyr]. 
Flux scales for the template spectrum are 
equally gridded in the logarithmic scale 
from $10^{10.5}$ to $10^{13.5}$, roughly corresponding to the range of 
stellar mass in \solarmass\ unit. 
Eleven stellar ages are considered; $t_{\rm{age}}=
0.1, 0.2, 0.5, 1.0, 1.5, 2.0, 2.5, 3.0, 3.5$, and $4.0$~Gyr 
which is younger than the age of the universe at $z=\zclq$. 
Metallicity $\log(Z/Z_\odot)$ is fixed to be a solar value (0) 
for the MIST isochrone. 

\begin{figure*}
 \begin{center}
    \includegraphics[width=175mm,bb=0 0 864 864]{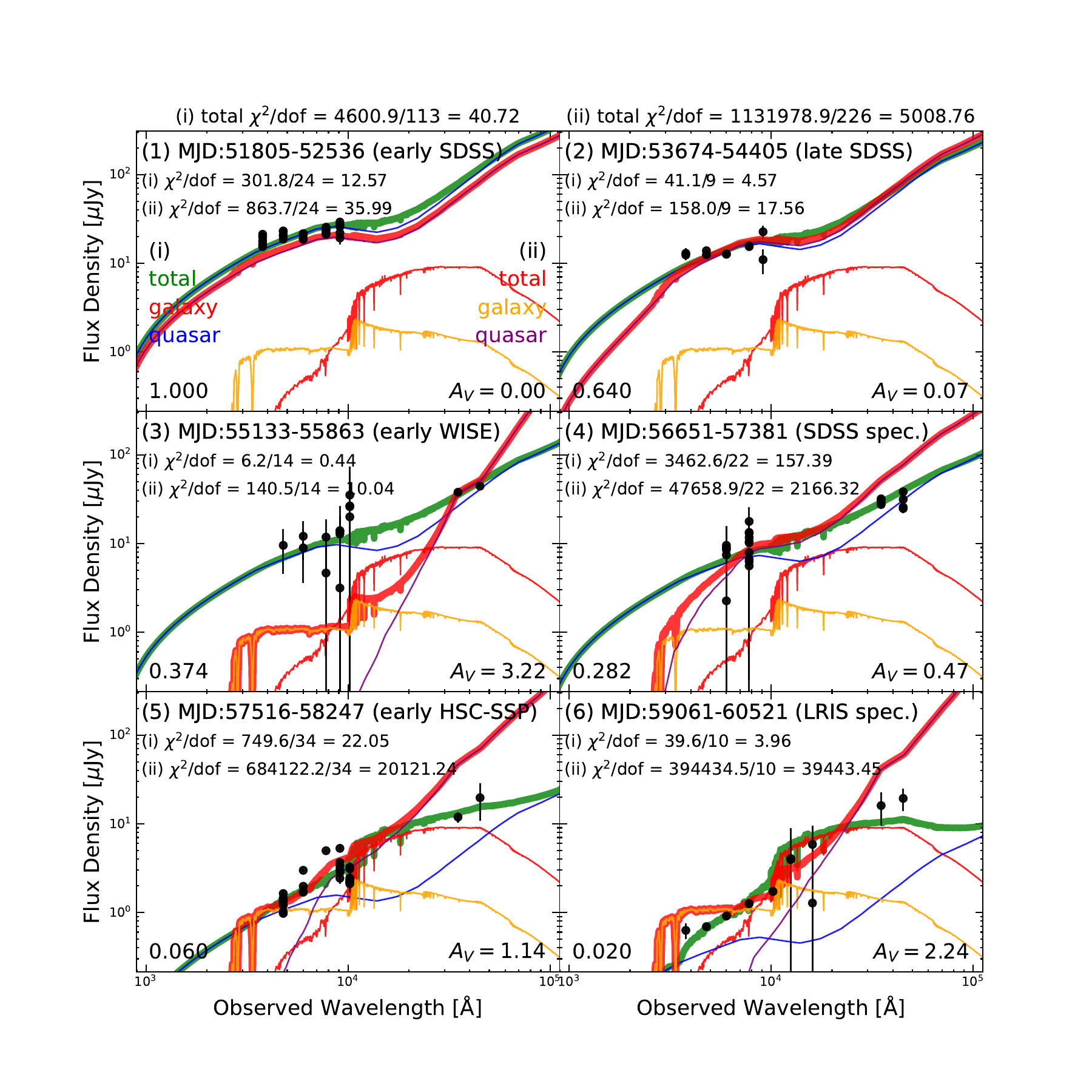}
 \end{center}
\caption{Time evolution of the 6-epoch SED. 
For the scenario 1), the total, AGN component, and galaxy component are shown in 
green, blue, and red, 
respectively. 
For the scenario 2), those are shown in 
light-green, purple, and orange, 
respectively. 
The black points are observed data. 
Reduced $\chi^2$ values for the two scenarios in each epoch and all the six epochs 
are shown in each panel and the top of the figure, respectively.
Best-fit relative scale factors for the AGN component in the scenario (i) and 
dust extinction $A_V$ in the scenario (ii) are also described at the bottom of each panel. 
{Alt text: 
Six-panel plot showing SEDs at six different epochs. Observed flux densities are plotted over wavelength. Model components representing total, galaxy, and quasar emissions from the two scenarios are also shown.
}
}\label{fig:figinpaper_sedevo}
\end{figure*}

We perform Bayesian SED fitting with a Markov Chain Monte Carlo (MCMC) 
approach for the six-epoch SEDs under the two scenarios.
The fitting is done simultaneously across all the six epochs, 
assuming significantly varying AGN flux scaling in the scenario (i), 
and no AGN flux variation but strong time-variable dust extinction in the scenario (ii). 
The fitting results are shown in Figure~\ref{fig:figinpaper_sedevo}. 
In the best-fit models for each scenario, 
the AGN flux scaling varies by a factor of 
$\sim\declinefactoreddratenew (=\releddratefirst/\releddratelast)$ 
over time in the scenario (i), 
while the dust extinction $A_V$ toward the AGN 
ranges from $\sim0$ to 
$\sim\avvaluelast$~mag 
in the scenario (ii). 

To evaluate which scenario better explains the data, 
we compute the Akaike Information Criterion (AIC) and Bayesian Information Criterion (BIC). 
The AIC values are $14,851$ for the scenario (i) and $18,401$ for the scenario (ii), 
resulting in $\Delta{AIC}=3,550$. Likewise, the BIC values are $14,879$ and $18,433$, respectively, 
giving $\Delta{BIC}=3,554$. 
Then, both the AIC and BIC favor the scenario (i), where AGN variability alone explains 
the six-epoch SEDs without invoking dust extinction. 
These large differences indicate strong support for the scenario (i) over the scenario (ii). 
The middle and lower-left panels of 
Figure~\ref{fig:figinpaper_sedevo} 
clearly highlight the advantage of scenario (i).
It indicates that the observed large decline in flux 
is attributed to the intrinsic decline in accretion rate. 
While the AGN component dominates the total SED 
in 2000's (in the SDSS imaging epochs (1) and (2)), 
the current SED is dominated by either of AGN and galaxy components 
depending on wavelength; 
rest-frame UV is still dominated by the AGN 
while rest-frame optical and NIR are dominated 
by the host galaxy 
(Figure~\ref{fig:figinpaper_sedevo}). 
This is also supperted by the latest MOIRCS $YJH$-band images, 
showing extended morphology of this object;  
non-zero ellipticity and larger FWHMs than point sources in the images. 
Note that a rapid disappearance of an inner part of the accretion disk 
within a certain radius, 
similar to a transitioning phase of X-ray binaries (JED-SAD) 
with a jet-emitting disk (JED) and a standard accretion disk (SAD) 
\citep{Marcel:2022aa}, 
would have a qualitatively similar effect on the spectral shape 
to the scenario (ii), where the flux density in shorter wavelengths 
selectively becomes smaller. 

Derived values from the fitting are 
stellar mass of \stellarmass\ [\solarmass], 
stellar age of 3.5~Gyr, 
and 
\sfrate\ SFR. 
The BH-stellar mass ratio of the quasar is 
$\sim1\times10^{-3}$, 
which is consistent 
with the local value \citep{Shankar:2019aa} 
although the BH mass of the quasar is highly uncertain 
as we describe in the previous section. 
These values are compared with those 
for main-sequence (MS) star-forming galaxies \citep{Speagle:2014aa} 
at $z\sim1.8$. 
SFR at $z\sim1.8$ of an MS galaxy with a stellar mass of 
\stellarmass [M$_\odot$] 
is 
$\sim160$ 
[M$_\odot$ yr$^{-1}$]. 
Although the obtained zero SFR of the AGN 
could be due to the assumption for SFH of the $\tau$-model, 
the SFR of the host galaxy is well below it, 
indicating that the host galaxy is a quiescent galaxy 
rather than a starburst galaxy.

\begin{figure}
 \begin{center}
    \includegraphics[width=82.95mm,bb= 0 0 458 500]{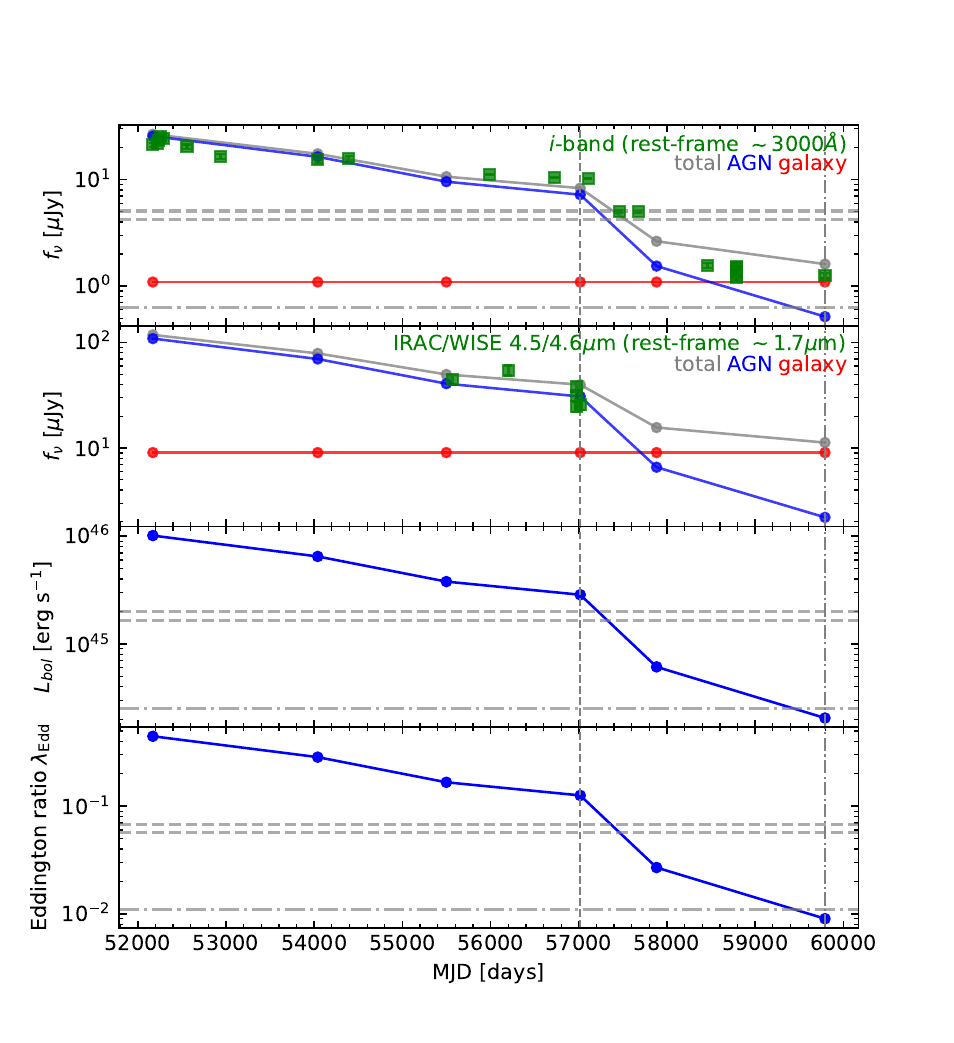}
 \end{center}
\caption{
Temporal changes of 
the apparent fluxes in $i$-band (top), 
the apparent fluxes in 4.5/4.6$\mu$m-band (second), 
bolometric luminosity (third), 
and 
Eddington ratios (bottom) 
are shown. 
The green squares are the observed brightness
while the gray, red, and blue circles 
are from the SED fitting 
for the total, AGN, and galaxy 
components, respectively,  
in the top and second panels. 
The $i$-band and 4.5/4.6$\mu$m-band 
are roughly corresponding to the rest-frame 0.3$\mu$m and $1.7\mu$m. 
The measurements in \citet{rakshit2020} 
and \citet{Wu:2022aa} 
are shown as gray dashed horizontal lines 
at the SDSS spectroscopic epoch as indicated 
gray dashed vertical lines. 
{Alt text: 
Four-panel plot showing temporal changes in the observed apparent fluxes 
in $i$-band and $4.5/4.6$~$\mu$m bands, bolometric luminosity, and Eddington ratio. 
In the apparent flux panels, observational data points, AGN components, and galaxy components estimated from the SED fitting are separately indicated.
}
}
\label{fig:figinpaper_bolLEddRatio}
\end{figure}

Figure~\ref{fig:figinpaper_bolLEddRatio} shows time evolution 
of obtained values from the SED fitting; 
observed $i$-band and 
4.5/4.6 $\mu$m flux densities, 
roughly corresponding to 
$\sim0.3~\mu$m\ and $\sim1.7~\mu$m\ 
in the rest-frame at $z=\zclq$, 
bolometric luminosity $L_{\rm{bol}}$, 
and 
Eddington ratios by assuming 
the black hole mass of 
$\log$\bhmass$=8.2$. 
As the AGN component luminosity decreases 
from the 1st epoch to the 6th epoch of the SED fitting 
(Figures~\ref{fig:figinpaper_sedevo} and \ref{fig:figinpaper_bolLEddRatio}), 
the bolometric luminosity and Eddington ratio also 
decrease by a factor of 
$\sim\declinefactoragnnew$ 
(Figure~\ref{fig:figinpaper_bolLEddRatio}). 
The Eddington ratio at the faintest phase is as low as 
$\sim\eddratiolownew$, 
decreased 
from the Eddington ratio of 
$\sim\eddratiohighnewimg$ 
in the SDSS 
early imaging epochs 
and $\sim\eddratiohighnewsdssspec$ 
in the SDSS spectroscopic epochs 
in 2000s (Figure~\ref{fig:figinpaper_bolLEddRatio}). 
This may indicate that 
the current accretion mode could not 
be standard accretion disk \citep{shakura1973} anymore. 
Note that the current SED fitting assumes 
a quasar-like SED and 
more accurately a different RIAF-like SED 
may better be considered in the faint phases in the SED fitting. 
The SEDs of accretion disks or flows and dusty torii are 
thought to change 
around Eddington ratios of 0.003 \citep{Gonzalez-Martin:2017aa} or 0.01 \citep{Abramowicz:1995aa,Sobolewska:2011aa}. 
We also note that the observed Eddington ratio change steps across 
the Eddington ratio threshold of a few percent 
where the accretion disk state transition happens 
proposed by \citet{Noda:2018aa}. 
At the Eddington ratios observed in the faint phase of this AGN, 
standard and ADAF-like disks could 
coexist \citep{Gonzalez-Martin:2017aa,Marcel:2022aa}. 

In the $\sim2$-year time ranges for each epoch in our SED fitting, 
any time delays of C\,{\sc iv} emission lines or NIR emission 
from the inner part of dusty torus 
relative to UV-optical continuum from the accretion disk 
are not necessary to be considered 
because the delays are shorter than the time range considered here; 
a few tens days for the C\,{\sc iv} emission lines \citep{Grier:2019aa} 
and $\sim100$~days for the NIR torus emission \citep{Minezaki:2019aa}. 
Note that rest-frame MIR (e.g., $>5~\mu$m) emission from the outer part 
of the torus is expected to have longer time delay 
than $\sim1$~year time scales.
Because of the non-detection in the rest-frame MIR for this object 
with longer-wavelength bands of WISE and Spitzer/MIPS, 
such remaining MIR torus emission cannot be investigated in this paper. 

\section{Timescales of Variability for J0218-0036}
\label{sec:secinpaper_timescalevari}

We compare the observed timescales of the large fading variability 
with theoretically expected timescales. 
As seen in Figure~\ref{fig:figinpaper_lc}, timescales 
in the exponential decay in flux density ($e$-folding timescales) 
are almost consistently $\sim2,000$~days, corresponding to 
$\sim700$~days in the quasar rest frame. 

We first consider 
timescales of accretion rate changes follow viscous \citep{Krolik:1999aa} 
or ``radial inflow'' timescales 
\tinfl\ 
as discussed in 
previous papers \citep{lamassa2015,macleod2016}. 
The timescale \tinfl\ 
is defined as 
\tinfl$=31\left(\frac{\alpha}{0.1}\right)^{-1}\left(\frac{\lambda_{\rm{Edd}}}{0.03}\right)^{-2}\left(\frac{\eta}{0.1}\right)^2\left(\frac{r}{10r_{g}}\right)^{3.5}\left(\frac{M_{8}}{1.7}\right)$ years 
where 
$\alpha$ is the ``viscosity'' parameter, 
$\eta$ is the efficiency of converting potential energy to radiation, 
$r_{g}$ is the gravitational radius $(=GM/c^{2})$, 
and 
$M_{8}=M_{\rm{BH}}/10^{8}M_{\odot}$. 
For example, a rest-frame wavelength of $\lambda_{\rm{rest}} = 3000\,\text{\AA} = 0.3\,\mu\mathrm{m}$
corresponds to an effective temperature of $T_{\rm{eff}} \approx 9.7 \times 10^3\,\mathrm{K}$,
which is typically reached at a radius of a few hundred gravitational radii ($r_g$)
in the Shakura--Sunyaev standard accretion disk.
Then, the timescale \tinfl would be 
as long as $10^5$~years
if $\alpha$ of 0.1 
and $\eta$ of $0.1$ 
are assumed. 
This is much longer than that observed for the quasar. 

Crossing time of dust 
cloud passing the broad line region and 
UV-optical emitting region 
$t_{\rm{cross,dust}}$ would be 
$t_{\rm{cross,dust}}\sim24M_{8}^{-1/2}L_{44}^{3/4}$~years \citep{macleod2016} 
where $L_{44}$ is monochromatic luminosity 
at 5100\AA\ divided by $10^{44}$~erg~s$^{-1}$ ($L_{5100}/10^{44}$~erg~s$^{-1}$), 
roughly corresponding to 
an order of 10-100~years depending on Eddington ratios 
\citep{Nenkova:2008aa,Nenkova:2008ab,Elitzur:2009wj,Elitzur:2012aa,macleod2016}. 
This is more consistent with the observations 
than the viscous timescale calculated above. 
However, 
the SED temporal changes are not likely to be attributed to 
time variability of dust extinction. 

Longer-term variability investigated using 20th-century data in optical 
and X-ray (Figure~\ref{fig:figinpaper_xrayplatelc}) 
would be interesting to be examined 
to explore the origin of the observed large brightness decline. 
There is a positive correlation between the soft X-ray and UV luminosities 
in AGN \citep{Lusso:2010aa}, usually connected via a parameter $\alpha_{\rm{OX}}$. 
For example, using the equation in \citet{Lusso:2010aa} 
on the relation between luminosities in 0.5-2~keV 
and 2500 \AA\ ($\sim7000$ \AA\ in the observed frame, 
roughly corresponding to $r$-band), 
the X-ray flux can be estimated as below. 
The extrapolated soft X-ray fluxes converted from observed $r$-band fluxes 
are shown in Figure~\ref{fig:figinpaper_xrayplatelc}. 
This is well consistent with the upper limits obtained for the ROSAT and RXTE data 
(\S\ref{sec:secinpaper_xraydata}). 
Unfortunately, 
no significant detections for this quasar 
are obtained in the 20th century data including the DSS and X-ray satellites.

In addition, 
spectral changes would give a key to understanding 
what is happening in the quasar, for example, 
an onset of outflow/inflow accompanied with the drastic fading phenomenon, 
by comparing the SDSS and LRIS spectra. 
We measured velocity shifts 
of the C\,{\sc iv} broad emission lines 
relative to the Mg\,{\sc ii} broad lines.  
Unfortunately, the Mg\,{\sc ii} lines are not well measured 
because of the low S/N ratios in both the spectra 
and no convincing results are obtained 
for the velocity shifts. 

\section{Summary}
\label{sec:secinpaper_summary}

We detected the dramatic apparent 
optical fading of the quasar at $z=\zclq$ 
by a factor of $\sim\declinefactorobservednew$ in optical 
over a period of $\sim20$~years 
by comparing the SDSS and HSC photometry of $\numofquasarswithinhscfootprints$ quasars spectroscopically identified in the SDSS projects. We further conducted follow-up optical imaging and spectroscopic and near-infrared imaging observations with $>4$m-class telescopes including Subaru, GTC, Keck, and SOAR telescopes. We combined the new data with the archival data and examined the temporal variability behavior over $\sim20$~years in detail and even the longer trend of the variability over $\sim70$~years in the observed frame. Our findings are summarized; 
(i) the brightness decline of the AGN component is 
a factor of $\sim\declinefactoragnnew$ 
from early 2000s to 2023 
and 
(ii) the observed brightness decline is attributed to a large decline in accretion rate rather than time-varying line-of-sight dust obscuration. These findings were concluded by the multi-component (time-varying AGN + constant galaxy) SED 
fitting over multi-epochs, which is well consistent 
with the optical spectra. The Eddington ratio decreased by 
by a factor of $\sim\declinefactoragnnew$ 
from $\sim\eddratiohighnew$ to $\sim\eddratiolownew$, 
although the black hole mass is highly uncertain because of the large variability nature of the quasar. The total brightness as of 2023 is dominated by the host galaxy rather than the AGN 
in the rest-frame optical while the AGN component is as comparable as the host galaxy in the rest-frame UV. 


\begin{ack}
%

These collaborative observations have become possible thanks to the
leadership of
the National Astronomical Observatory of Japan (NAOJ) and
the Instituto de Astrofísica de Canarias (IAC).
This research is 
based in part on data collected at Subaru Telescope,
which is operated by NAOJ and
in part on observations made with the GTC telescope,
in the Spanish Observatorio del Roque de los Muchachos of IAC,
under Director’s Discretionary Time.
We are honored and grateful for the opportunity of observing
the Universe from Maunakea, which has the cultural, historical and
natural significance in Hawaii.

JBG and JAP acknowledges support from the Agencia Estatal de Investigaci\'{o}n 
del Ministerio de Ciencia, Innovaci\'{o}n y Universidades (MCIU/AEI) under 
grant PARTICIPACIÓN DEL IAC EN EL EXPERIMENTO AMS and the European Regional 
Development Fund (ERDF) with 
reference PID2022-137810NB-C22. 
JBG and JAAP acknowledges the support from 
the Severo Ochoa Centres Excellence Programme 2020-2024 (CEX2019-000920-S).

This research is also 
based on observations obtained with MegaPrime/MegaCam, a joint project of CFHT and CEA/DAPNIA, at the Canada-France-Hawaii Telescope (CFHT) which is operated by the National Research Council (NRC) of Canada, the Institut National des Science de l'Univers of the Centre National de la Recherche Scientifique (CNRS) of France, and the University of Hawaii. The observations at the Canada-France-Hawaii Telescope were performed with care and respect from the summit of Maunakea which is a significant cultural and historic site.
This research has made use of the NASA/IPAC Infrared Science Archive, which is funded by the National Aeronautics and Space Administration and operated by the California Institute of Technology. 
The National Geographic Society - Palomar Observatory Sky Atlas (POSS-I) was made by the California Institute of Technology with grants from the National Geographic Society.
The Second Palomar Observatory Sky Survey (POSS-II) was made by the California Institute of Technology with funds from the National Science Foundation, the National Aeronautics and Space Administration, the National Geographic Society, the Sloan Foundation, the Samuel Oschin Foundation, and the Eastman Kodak Corporation. The Oschin Schmidt Telescope is operated by the California Institute of Technology and Palomar Observatory.
This work is based in part on observations obtained with \textit{XMM-Newton}, an ESA science mission with instruments and contributions directly funded by ESA Member States and NASA. It also makes use of data obtained from the \textit{ROSAT} mission, a cooperative project of the Max-Planck-Institut f\"ur extraterrestrische Physik (MPE) and NASA, and from the \textit{Rossi X-ray Timing Explorer} (RXTE), provided via the High Energy Astrophysics Science Archive Research Center (HEASARC) at NASA's Goddard Space Flight Center.

This work made use of Astropy:\footnote{http://www.astropy.org} a community-developed core Python package and an ecosystem of tools and resources for astronomy \citep{Astropy-Collaboration:2013aa,Astropy-Collaboration:2018aa,Astropy-Collaboration:2022aa}. 

\end{ack}

\section*{Funding}
This research was supported by 
JSPS KAKENHI Grant numbers 
16H01088, 
20H00179, 
21H00066, 
24H00027 (T.M.), 
17K05389, 
25K07370 (T.K.), 
JP22H01266, and JP23K22537 (Y.T.). 
The grants from Yamada Science Foundation and the Sumitomo Foundation (2200605) 
also supports this study (T.K.).

\section*{Data availability} 
The data underlying this article will be shared on reasonable request to the corresponding author.

\bibliography{pasj}

\end{document}